\documentclass[printer]{aa}
\usepackage{natbib}
\usepackage{amssymb}
\usepackage{graphicx}
\usepackage{hyperref}
\usepackage{comment}
\usepackage[caption=false]{subfig}
\newcommand{\Mdisc}{\mbox{$M_\mathrm{disc}$}}

\newcommand{\Rdisc}{\mbox{$R_\mathrm{disc}$}}

\newcommand{\Msun}{\mbox{$\mathrm{M}_{\odot}$}}

\newcommand{\dMdtdisc}{\mbox{$\dot{M}_{\mathrm{disc}}$}}
\newcommand{\dMdtstar}{\mbox{$\dot{M}_{\mathrm{star}}$}}
\newcommand{\dMdtISM}{\mbox{$\dot{M}_{\mathrm{ISM}}$}}
\newcommand{\dMdtstrip}{\mbox{$\dot{M}_{\mathrm{strip}}$}}

\makeatletter
\newcommand\footnoteref[1]{\protected@xdef\@thefnmark{\ref{#1}}\@footnotemark}
\makeatother

\begin{document}

\title{Face-on accretion onto a protoplanetary disc}
\titlerunning{Accretion onto a protoplanetary disc}
\author{T.P.G. Wijnen\inst{1,2}, O.R. Pols\inst{1}, F.I. Pelupessy\inst{2,3}, S. Portegies Zwart\inst{2}}
\authorrunning{T.P.G. Wijnen et al.}
\institute{Department of Astrophysics/IMAPP, Radboud University Nijmegen, P.O. Box 9010, 6500 GL Nijmegen, The Netherlands\\
\email{thomas.wijnen@astro.ru.nl}
\and Leiden Observatory, Leiden University, PO Box 9513, 2300 RA Leiden, The Netherlands
\and Institute for Marine and Atmospheric research Utrecht, Utrecht University,
Princetonplein 5, 3584 CC Utrecht, The Netherlands
\offprints{T.P.G. Wijnen}
}
\date{Received ..../ Accepted ....}

\abstract{Stars are generally born in clustered stellar environments, which can affect their subsequent evolution. An example of this environmental influence can be found in globular clusters (GCs) harbouring multiple stellar populations. Bastian et al. recently suggested an evolutionary scenario in which a second (and possibly higher order) population is formed by the accretion of chemically enriched material onto the low-mass stars in the initial GC population to explain the multiple stellar populations. The idea, dubbed `Early disc accretion', is that the low-mass, pre-main sequence stars sweep up gas expelled by the more massive stars of the same generation into their protoplanetary disc as they move through the cluster core. The same process could also occur, to a lesser extent, in embedded stellar systems that are less dense.}
{Using assumptions that represent the (dynamical) conditions in a typical GC, we investigate whether a low-mass star of $0.4\,\Msun$ surrounded by a protoplanetary disc can accrete a sufficient amount of enriched material to account for the observed abundances in `second generation' GC stars. In particular, we focus on the gas loading rate onto the disc and star as well as on the lifetime and stability of the disc.}
{We perform simulations at multiple resolutions with two different smoothed particle hydrodynamics codes and compare the results. Each code uses a different implementation of the artificial viscosity.}
{We find that the gas loading rate is about a factor of two smaller than the rate based on geometric arguments, because the effective cross section of the disc is smaller than its surface area. Furthermore, the loading rate is consistent for both codes, irrespective of resolution. Although the disc gains mass in the high resolution runs, it loses angular momentum on a time scale of $10^4$ years. Two effects determine the loss of (specific) angular momentum in our simulations: 1) continuous ram pressure stripping and 2) accretion of material with no azimuthal angular momentum. Our study, as well as previous work, suggests that the former, dominant process is mainly caused by numerical rather than physical effects, while the latter is not. The latter process, as expected theoretically, causes the disc to become more compact and increases the surface density profile considerably at smaller radii.}
{\textnormal{The disc size is determined in the first place by the ram pressure exerted by the flow when it first hits the disc. Further evolution is governed by the decrease in the specific angular momentum of the disc as it accretes material with no azimuthal angular momentum. Even taking into account the uncertainties in our simulations and the result that the loading rate is within a factor two of a simple geometric estimate}, the size and lifetime of the disc are probably not sufficient to accrete the amount of mass required in the Early disc accretion scenario.}
%

\maketitle

\section{Introduction}

Stars generally form in clusters \citep{lada03}. These dense environments affect the formation and evolution of the stars they host. For example, globular clusters (GCs) were once considered the archetype of coeval, simple stellar populations, but during the last two decades they have been shown to harbour multiple stellar populations. Observations imply that a considerable fraction (up to 70\,\%) of the stars currently in GCs has a very different chemical composition from the initial population \citep[see e.g.][]{gratton12}. They indicate that a second (and in some cases even higher order, e.g. \citet{piotto07}) population\footnote{Most scenarios proposed to date imply subsequent epochs of star formation and hence refer to multiple \emph{generations} of stars in a GC. Since it is still not clear whether GCs can facilitate an extended star formation history or if the enriched stars actually belong to the initial population, we will refer to multiple \emph{populations} of stars.} of stars has formed from material enriched by ejecta from first generation stars. 

To explain the formation of these enriched stellar populations, several scenarios have been proposed \citep[see e.g.][]{decressin07-1, dercole08, de_mink09, bastian13-2}. One of the recently proposed scenarios applies in particular to star formation in GCs, but could, in theory, also leave its signature on stellar systems that are less dense. \citet{bastian13-2}, hereafter B13, have suggested a scenario in which the enriched population is not formed by a second star formation event, but rather by the accretion of enriched material, that was expelled by the high-mass stars of the initial population, onto the low-mass stars of the same (initial) population. Because Bondi-Hoyle accretion, i.e. gravitational focusing of material onto the star, is unlikely to be efficient in a GC environment with a high velocity dispersion, they suggest that the protoplanetary discs of low-mass stars sweep up the enriched matter. In order to account for the observed abundances in the enriched population, the low-mass stars have to accrete of the order of their own mass, i.e. a 0.25 $\Msun$ star has to accrete about 0.25 $\Msun$ of enriched material in the most extreme case (as inferred from, e.g., the main sequence of NGC2808 \citep{piotto07}). The time scale of this scenario is limited by the lifetimes and sizes of protoplanetary discs. B13 assume that the protoplanetary discs can accrete material for up to 20 Myr. Current disc lifetimes are observed to be 5-15 Myr, but B13 argue that their lifetimes may have been considerably longer in GCs. The accretion rate averaged over 20 Myr therefore has to be about $10^{-8} \Msun / yr$. In their scenario, they assume that the accretion rate is proportional to the size of the disc, $\pi R_{\rm disc}^2$, density of the ISM, $\rho_{\rm ISM}$, and the velocity, $v_{\rm ISM}$, of the disc with respect to the ISM, i.e. $\dot{M} \propto \rho_{\rm ISM} v_{\rm ISM} \pi R_{\rm disc}^2$. Furthermore, they assume an average and constant disc radius of 100 AU. In this work, we test several of these assumptions of the early disc accretion scenario.

A similar scenario has been studied before by \citet{moeckel09}, M09 hereafter. They performed smoothed particle hydrodynamics simulations of a protoplanetary disc that is embedded in a flow of interstellar medium (ISM) with a velocity of $3\,\mathrm{km\, s^{-1}}$. They found that the mean accretion rate onto the star equals the rate expected from Bondi-Hoyle theory, whether a disc is present or not. We note that for the parameters they assumed, the theoretical Bondi-Hoyle radius is almost twice the radius of the disc.
Here we follow up on the work by M09 by simulating the accretion process onto a protoplanetary disc for the typical conditions expected in a GC environment. We directly compare the outcome of two different smoothed particle hydrodynamics codes for the same set of initial conditions and different particle resolutions. We first discuss both the physical and numerical effects in our reference model and subsequently we compare the results of the different codes and particle numbers.

\section{Expected physical effects}

In dense stellar systems, where both the velocity dispersion and the possibility of close stellar encounters are high, we expect the following physical effects to play an important role in the process of accretion onto protoplanetary discs: ram pressure, angular momentum transport, dynamical encounters and external photo-evaporation. 
 
\subsection{Ram pressure stripping}
\label{sec:rampressure_eda1}
As the protoplanetary disc moves through the ISM, it experiences ram pressure, $P_{\rm ram} = \rho_{\rm ISM} v_{\rm ISM}^2$. This drag force can truncate the disc, depending on the gravitational force of the disc that keeps the latter bound to the star. By equating the gravitational force per unit area, or `pressure', $P_{\rm grav}=GM_*\Sigma(r)r^{-2}$, of the disc to $P_{\rm ram}$, we determine beyond which radius the ram pressure dominates and the disc is expected to be truncated. This truncation radius is given by:
\begin{equation}\label{eq:ramradius_eda1}
R_{\rm trunc} = \left(\frac{GM_*\Sigma_0 \mathbf{r_0^n}}{\rho_{\rm ISM} v_{\rm ISM}^2}\right)^{\frac{1}{n+2}}
\end{equation}
with $M_*$ the mass of the star and we have assumed that the surface density profile of the disc can be written as $\Sigma(r)=\Sigma_0 (r/r_0)^{-n}$, where $r_0$ is an arbitrary but constant radius to which $\Sigma_0$ is scaled. After the material at radii larger than $R_{\rm trunc}$ has been stripped from the disc, we expect further evolution of the disc to be determined by the redistribution of angular momentum due to accretion and viscous evolution. The pressure in the mid-plane of the disc is at least one order of magnitude smaller than the gravitational `pressure' and therefore does not play a significant role in resisting the ram pressure.

\subsection{Redistribution of angular momentum}
\label{sec:am_redistribution_eda1}
The redistribution of angular momentum in the disc is governed by two processes:(1) the accretion of material with no angular momentum material with respect to the disc and (2) the viscous evolution of the disc and consequent redistribution of its mass and angular momentum. 
\subsubsection{Accretion of ISM}
The accretion of material with no azimuthal angular momentum lowers the specific angular momentum of the disc. Since the total angular momentum of the disc has to be conserved, mass and angular momentum will be redistributed within the disc. We can estimate this redistribution to first order, if we consider the disc to consist of concentric rings with thickness $dr$ and mass $m_{\rm ring}(r) = 2 \pi r \Sigma(r) dr$. In a time interval $dt$, the ring will accrete an amount of mass from the ISM equal to $m_{\rm accr}(r) = 2 \pi r \rho_{\rm ISM} v_{\rm ISM} dt dr$. The specific angular momentum, $h$, in the ring will decrease by a factor $\Sigma(r)/(\Sigma(r) + \rho_{\rm ISM} v_{\rm ISM} dt)$ and the ring will migrate to a smaller radius that corresponds to its new specific angular momentum. This process causes inward migration of material that belongs to the disc and leads to contraction of the disc. This derivation does not take into account interaction between adjacent rings, which is determined by viscous processes.

\subsubsection{Viscous redistribution}

Although the nature of the viscous processes that occur in accretion discs are still debated \citep[see e.g.][]{armitage11}, we do know that they are responsible for transporting material inwards through the disc.
When this happens, some material has to move outward to conserve the total angular momentum of the disc, $J_{\rm disc}$. Both viscous evolution and accretion of ISM cause material to drift inwards where it is eventually accreted onto the star and lost from the disc together with the angular momentum it carried. The outward spreading of material at the outer edge of the disc could make it more vulnerable to ram pressure stripping, which in turn also robs the disc of its angular momentum.

\subsection{Dynamical encounters}

In dense stellar systems, disc radii have been shown to be truncated due to close stellar encounters \citep{breslau14, rosotti14, vincke15}. The last study shows that in dense clusters ($\bar{\rho}_{\rm cluster} \approx 500\,\mathrm{pc^{-3}}$), almost 40\,\% of the discs is smaller than 100 AU and the median disc radius can be as small as 20 AU in the core. Close stellar encounters thus affect the surface area of the disc and thereby also the rate at which the disc can sweep up ISM. In this work we only focus on the hydrodynamic aspects of accretion onto protoplanetary discs, in particular the accretion rate. The disc radius we find in our simulations is of similar size as the 20 AU found by \citet{vincke15}. The question whether the process of ram pressure or dynamical encounters dominates the truncation of the disc in embedded dense stellar systems is beyond the scope of this paper.

\subsection{External photo-evaporation}

Globular clusters host a large number of massive stars in the early phases of their evolution and the large UV flux they produce may also strip material from the disc. Studies have shown that the fraction of stars that have discs can decrease by a factor of two close to O stars \citep[see e.g.][]{balog07, guarcello07, guarcello09, fang12}, but \citet{richert15} argue that these results could be partly affected by sample incompleteness. Recently, \citet{facchini16} estimated that the mass loss rate from the outer edge of a protoplanetary disc due to photo-evaporative winds can be of the order of $10^{-8} - 10^{-7}\, \Msun / \mathrm{yr}$, see their Fig. 12. We do not take radiative processes into account in this work but we note that any mass loss from the disc, additional to that found in this work, will shorten its lifetime with respect to our findings.

\section[]{Numerical set-up}
\label{sec:set-up_eda1}

\begin{figure}
\centering
    \includegraphics[width=0.49\textwidth]{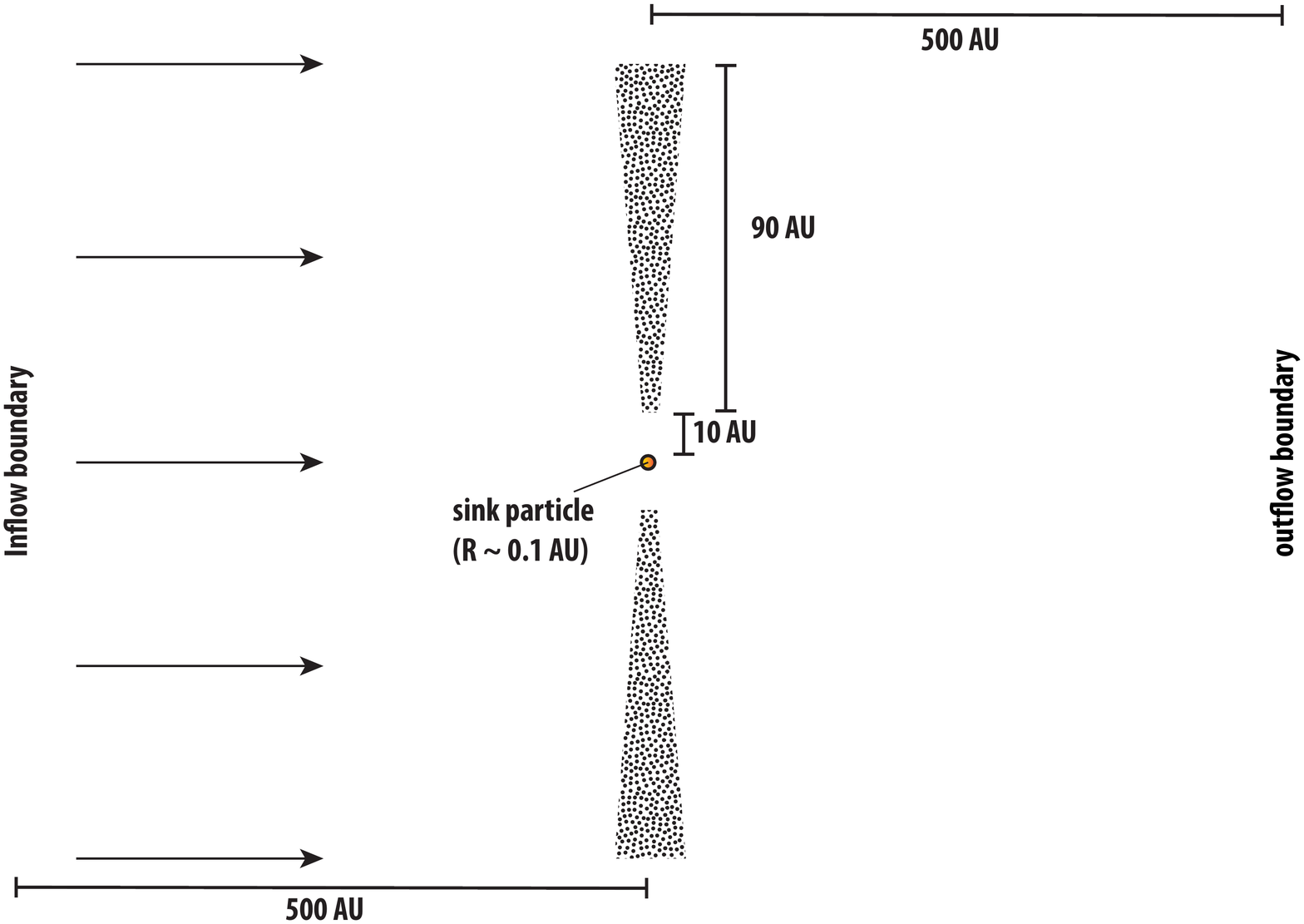}
    \caption{A schematic overview of the numerical set-up.\label{fig:schematics_eda1}}  
\end{figure}

Our simulations are performed using the AMUSE environment \citep{portegies_zwart13, pelupessy13}\footnote{\url{http://www.amusecode.org}}. AMUSE is a Python framework in which one can use and combine a variety of astrophysical codes that have been published and well tested by the community, i.e. `community codes'. The simulation is set up in Python and AMUSE takes care of the communication between Python and the code(s) that one wishes to use. This way, it is very easy to use the same set-up with different community codes and compare their outcome. 

Currently, two different smoothed particle hydrodynamics (SPH) codes are included in AMUSE, i.e. Fi \citep{pelupessy04} and Gadget2 \citep{springel05,springel01}. These two codes solve the dynamics of a self-gravitating hydrodynamic fluid using a Tree gravity solver \citep{hernquist89}. In this work, we use Gadget2 for our reference model and we verify the consistency and robustness of our results by comparing with Fi. Both SPH codes include self-gravity.

Because we want to be able to perform future simulations for a wide range of parameters, we need to find a balance between computational effort and convergence, i.e. the most efficient set-up. In order to test whether our particle resolution is sufficient, we perform simulations with different numbers of particles in the disc. This way, we can test whether the results converge and the numerical noise diminishes. The verification and validation will be discussed in Sec. \ref{sec:convergence_eda1}.  

For our reference model, we use the vanilla version of Gadget2 as it is freely available online\footnote{\url{http://www.mpa-garching.mpg.de/gadget}} and included in AMUSE, with the only exception that we have implemented the \citet{morris97} viscosity formalism as is done in M09, see Sec. \ref{sec:visc_eda1}. 

The set-up of our simulations is as follows (see Fig. \ref{fig:schematics_eda1}): a stationary protoplanetary disc is positioned coaxially in a cylindrical, laminar flow of gas, representing the ISM. The protoplanetary disc is positioned in the centre of the cylinder. The inflow and outflow boundaries are each located at a distance of 500 AU from the centre of the protoplanetary disc. The radius of the cylinder is also set to 500 AU. Particles that flow outside the computational domain are removed from the simulation. The inflow consists of new particles. The parameters we have adopted are listed in Table \ref{tb:parameters_eda1}. 

\begin{table}
\begin{center}
\caption{Initial values of the parameters in our simulations. The parameters are the same for every simulation, except for the number of disc particles which may vary from simulation to simulation.}
\label{tb:parameters_eda1}
\begin{tabular}{lrl}
\hline
\textbf{Parameter}&\textbf{Value}&\textbf{Description}\\
\hline
$n$&$5 \times 10^6 \, \mathrm{cm^{-3}}$&Number density of ISM\\
$v_{\rm ISM}$&$20\,\mathrm{km\, s^{-1}}$&\\
$\mu$&2.3&Mean molecular weight\\
$M_*$&$0.4\,\Msun$&\\
$M_{\rm disc}$&$0.004\,\Msun$&\\
$R_{\rm disc, inner}$&10\,AU&\\
$R_{\rm disc, outer}$&100\,AU&\\
EoS&Isothermal&Equation of state\\
$T$&25\,K&Temperature of gas particles\\
$c_s$&$0.3\,\mathrm{km\, s^{-1}}$&Sound speed\\
$R_{\rm sink}$&0.09\,AU&Sink particle radius\\
$N_{\rm neighbours}$& $64\pm2$&\\
$\epsilon_{\rm grav}$&$10^{-2}\,\mathrm{AU}$&Gravitational softening length\\
$\alpha_{\rm SPH}$&0.1&Artificial viscosity parameter\\
\hline
$\mathrm{N_{\rm disc}}$& $4000-128000$&Number of disc particles\\
\hline
\end{tabular}
\medskip
\end{center}
\end{table}

\subsection{Initial conditions and parameters}

We adopt equal mass particles for the flow and the disc in order to prevent spurious forces on the interface between the ISM and the disc. We set the number of neighbours in the SPH codes to $64 \pm 2$ and use an isothermal equation of state. The sound speed, $c_{\rm s}$, equals $0.3\,\mathrm{km\, s^{-1}}$, which corresponds to a temperature of approximately 25 K for all particles. The isothermal equation of state can be justified by considering the cooling time scale:
\begin{equation}\label{eq:cooling_tau_eda1}
\tau_{\rm cool} = \frac{\frac{3}{2}k_{B}T}{n \Lambda(n,T)}
\end{equation}
where $n$ is the number density, $k_{B}$ the Boltzmann constant, $T$ the temperature and $\Lambda(n,T)$ the cooling rate. For $T$ = 25K and $n \geq 10^3 \mathrm{cm}^{-3}$, the cooling time scale  $\tau_{\rm cool} \lesssim 16$ years, assuming $\Lambda \approx 10^{-26} \mathrm{erg \ cm^3 \ s^{-1}}$ \citep{neufeld95}. Any departure from the equilibrium temperature of 25K will therefore be restored quickly with respect to our simulation time scale. We discuss the parameters and assumptions for the ISM and the disc separately below in Sec. \ref{sec:ism_eda1} and \ref{sec:disc_eda1}, respectively.

\subsubsection{Interstellar Medium}
\label{sec:ism_eda1}
By adopting a temperature of 25 K, we assume that cooling of the ISM is much more efficient than the radiative transfer of heat from nearby stars. It has been suggested that the Lyman-Werner flux plays an important role in the formation of multiple populations \citep{conroy11}, but a survey on 130 young massive clusters have shown little to no ionized gas \citep{bastian13}. Young massive clusters are considered to be the modern-day counterpart of proto-GCs and this study therefore implies that heating does not play an important role in the environment where the accretion process is believed to take place.

We can also estimate the cooling time scale for stellar ejecta from Eq. \ref{eq:cooling_tau_eda1}, to justify that the expelled gas cools fast to low temperatures. As a lower limit for the density of stellar ejecta we assume $n = 10^2 \mathrm{cm}^{-3}$ and $T = 10^4$ K \citep[][ see also B13]{smith07}. Combined with an appropriate cooling rate of $\Lambda \approx 10^{-25} \mathrm{erg\,cm^3 / s}$ \citep{richings14} this results in a cooling time scale of roughly 6500 years. We consider this an upper limit, because assuming a higher density for the stellar ejecta would result in an even shorter cooling time scale. Consequently, the cooling time scale is at least three to four orders of magnitude smaller than the $10^{7}$ years time scale on which the early disc accretion scenario is expected to take place. Our assumption that the ISM has cooled sufficiently and can be approximated with a temperature of 25 K is therefore justified.

To estimate the density of the ISM in the early disc accretion scenario we use the values given in B13 for the available processed material and the core radius of a typical GC, respectively $1.3\times10^5\,\Msun$ and 1 parsec. The average density would then be around  $2 \times 10^{-18} \,\mathrm{g\,cm^{-3}}$. However, the density will vary and can locally be higher. To provide a better comparison with M09, we adopt their assumed number density of $n = 5 \times 10^{6}\,\mathrm{cm^{-3}}$ and mean molecular weight $\mu=2.3$, which translates to a mass density, $\rho_{\rm ISM}$, of $1.92 \times 10^{-17} \,\mathrm{g\,cm^{-3}}$. In the case of GCs with multiple stellar populations, $\mu$ may be somewhat larger because the enriched populations exhibit a helium enhancement compared to primordial molecular clouds. 

We assume an inflow velocity, $v_{\rm ISM}$, of 20 $\mathrm{km \ s}^{-1}$ in order to approximate the typical velocity dispersion in GCs. This means that our set-up is in the supersonic regime and the treatment of the artificial viscosity is important. We will discuss the artificial viscosity in Sec. \ref{sec:visc_eda1}. The high velocity gives a very small Bondi-Hoyle accretion radius, as we will discuss in Sec. \ref{sec:sink_eda1}. The inflow is modelled by adding a slice of ISM with thickness $v_{\rm ISM}dt$ at the inflow boundary and random uniformly distributing the SPH-particles within this slice. The ISM flow reaches the disc after $\approx$ 100 years and the computational domain contains  $\approx 6N_{\rm disc}$ ISM-particles when it is completely filled.

\subsubsection{Disc}
\label{sec:disc_eda1}
In the case of a disc in a steady state, the mid-plane temperature profile follows a simple power-law, $T_c \propto r^{-p}$, and the surface density profile, $\Sigma$, is proportional to $r^{p-\frac{3}{2}}$, assuming a constant viscosity parameter $\alpha_{\nu}$ in the \citet{shakura73} formalism \citep[see e.g.][]{armitage11}. Since we assume a constant temperature in the whole disc,  $p = 0$  and $\Sigma \propto r^{-\frac{3}{2}}$, which corresponds to a minimum mass solar nebula \citep{Weidenschilling77, hayashi81} and is also assumed in e.g. M09 and \citet{rosotti14}. The typical temperature of a protoplanetary disc at radii $> 10$ AU is roughly consistent with a temperature of 25 K \citep[$\approx$ 20K, see e.g.][]{armitage11}. Although heating of the outer layers of the disc may cause photoevaporation, the mid-plane of the disc is shielded. At approximately 10 AU, the disc has to be heated to temperatures $> 10^3$ K to effectively lose mass due to photoevaporation. An $0.4\,\Msun$ star has an effective temperature of $3\times 10^3$ K, which is not high enough to unbind gas from the surface layer of the disc at radii $> 10$ AU. At radii $< 10$ AU extreme-ultraviolet radiation from the star may cause mass loss from the disc in the order of $10^{-11}-10^{-10}\,\Msun/$yr \citep{armitage11, font04}. This is 2 to 3 orders of magnitude less than the mass loss rates found in this work and we conclude that taking heating by the star into account would not affect our results.

The stability of differentially rotating gaseous discs against self-gravity can be expressed in terms of the Toomre parameter $Q$ \citep{toomre64}, which is defined as:
\begin{equation}
Q = \frac{c_s \Omega}{\pi G \Sigma}
\end{equation}
where $\Omega$ is the angular frequency. A disc becomes unstable if $Q$ is less than unity at the outer edge of the disc. In our set-up, $Q$ has a value of $\approx 44$. Self-gravity is therefore not expected to lead to instabilities. As in M09, we assume that the initial mass of the disc is 1\% of the mass of the star and we set the outer radius of the disc to 100 AU and the inner radius to 10 AU. Realistically, the disc should probably extend inwards towards the radius of the star, but it is computationally very expensive to simulate the disc on the orbital time scales at these small radii. During the simulation, particles in the disc will migrate inwards due to viscous evolution of the disc and are accreted onto the star, resulting in a disc that extends further inwards. This set-up allows us to postpone the expensive calculations towards later times in our simulations, thereby significantly decreasing the duration of our simulations.

We position the disc perpendicular to the flow (see Fig. \ref{fig:schematics_eda1}). This perfect alignment may not occur frequently in nature, but allows to discern the relevant physical processes more clearly. In future work we will investigate the influence of giving the disc an inclination with respect to the flow direction.

\subsubsection{Star}
\label{sec:sink_eda1}
We assume that the mass of the star is concentrated in a single point, which has a mass of $0.4\,\Msun$. This corresponds to the typical mass expected in the early disc accretion scenario. The point mass is added as a collisionless particle to the SPH code and we treat it as a sink particle, i.e. it can accrete gas particles that fall within a certain (fixed) radius. The sink particle absorbs the mass and momentum carried by the gas particles it accretes.

We assume the radius of the sink particle is 5\% of the Bondi-Hoyle accretion radius, $R_{\rm A}$, defined as \citep[see also][]{Bondi52, shima85}:

\begin{equation}\label{eq:ra_eda1}
R_{\rm A} = \frac{2GM_{*}}{c_s^2 +v_{\rm ISM}^2}\\
\end{equation}
in which $M_{*}$ is the mass of the star, $c_s$ is the sound speed and $v_{\rm ISM}$ the velocity of the ISM (with respect to the star). Using the values mentioned above, gives us an accretion radius of 1.8 AU, which is significantly smaller than in M09, where $R_{\rm A} \approx 2 R_{\rm disc}$.

\subsection{Artificial viscosity}
\label{sec:visc_eda1}

Since the typical velocity dispersion in a GC is highly supersonic, it is important to resolve shocks. In order to do so, SPH codes introduce a numerical viscosity which is characterised by the parameters $\alpha_{\rm SPH}$ and $\beta_{\rm SPH}$, where $\beta_{\rm SPH}$ has been introduced to prevent particle interpenetration in shocks with high Mach numbers \citep{springel10}. Typical values for the parameters are $\alpha_{\rm SPH} \simeq 0.5-1$ and $\beta_{\rm SPH}=2\alpha_{\rm SPH}$. 

The numerical (shear) viscosity can also be used to model the physical viscous transport of matter in an accretion disc \citep{artymowicz94}, but this implies lower values of $\alpha_{\rm SPH}$. This value can be derived using \citet{artymowicz94} to relate the artificial viscosity parameter $\alpha_{\rm SPH}$ to the standard viscosity parameter $\alpha_{\nu}$ proposed by \citet{shakura73}. Assuming $\alpha_{\nu} \approx 0.01$ \citep{armitage11}, would correspond to an $\alpha_{\rm SPH}$ of roughly 0.02 in our reference model ($\alpha_{\rm SPH} \propto \sqrt[3]{N_{\rm disc}}$), which is too low for numerical reliability. We therefore set $\alpha_{\rm SPH}$ initially to 0.1, as was also done in e.g. M09 and \citet{rosotti14}. We have implemented the viscosity switch proposed by \citet{morris97} in Gadget2\footnote{This implementation is done in the source code of Gadget2, which is not according to the philosophy of AMUSE, but always possible if required.}. In this treatment of the viscosity every particle has its individual viscous parameter $\alpha_{\rm SPH}$ (and $\beta_{\rm SPH}=2\alpha_{\rm SPH}$), which is important because we simultaneously need a low value of $\alpha_{\rm SPH}$ in the disc and a high value of $\alpha_{\rm SPH}$, up to 1, to resolve the shock in front of the disc. In the SPH code Fi, the viscosity remains constant throughout the simulation at a value of $\alpha_{\rm SPH} = 0.1$. We have adopted $\beta_{\rm SPH}=1$ for Fi. With these values we minimize the viscous stresses in the disc, which due to the low relative velocities are dominated by the first order $\alpha_{\rm SPH}$ term, while preventing particle interpenetration in the strong accretion shock. We have tested lowering the $\beta_{\rm SPH}$ value below the adopted value and found that it can cause numerical artefacts.

Following \citet{artymowicz96}, we calculate the time scale on which the disc undergoes significant viscous evolution, $\tau_{\nu}=Re\,\Omega^{-1}$, assuming the Reynolds number, $Re$, and angular frequency, $\Omega$, are given by:
\begin{equation}\label{eq:reynolds_eda1}
Re = \frac{(r/\langle h \rangle)^2}{\alpha_{\nu}}\\
\Omega = \sqrt{\frac{GM_{*}}{r^3}}\\
\end{equation}
with $r$ the radial distance to the star and $\langle h \rangle$ the (resolution-dependent) average smoothing length in the disc at that radius. For a simulation of 16.000 disc particles, and assuming the corresponding $\alpha_{\nu}$, this gives us a viscous time scale, $\tau_{\nu}$, of roughly 10.000 years at 10 AU. We can also calculate the physical viscous time scale at 10 AU by replacing $\langle h \rangle$ with the scale height of the disc, $H=\frac{c_s}{\Omega}$, at that radius. This leads to a time scale of $3\times10^5$ years. In order to remain safely in the regime where the simulation is not dominated by viscous evolution of the disc, we evolve the whole set-up for 2500 years and start the inflow at $t=0$. The numerical viscosity in the disc changes during the simulations with Gadget2 and we will discuss the numerical effects in Sec. \ref{sec:convergence_eda1} and \ref{sec:consistency_eda1}.

\subsection{Distinguishing the disc from the ISM}
\label{sec:disc_recognition_eda1}
\begin{figure*}
\centering
    \subfloat[(a)]{\includegraphics[width=0.49\textwidth]{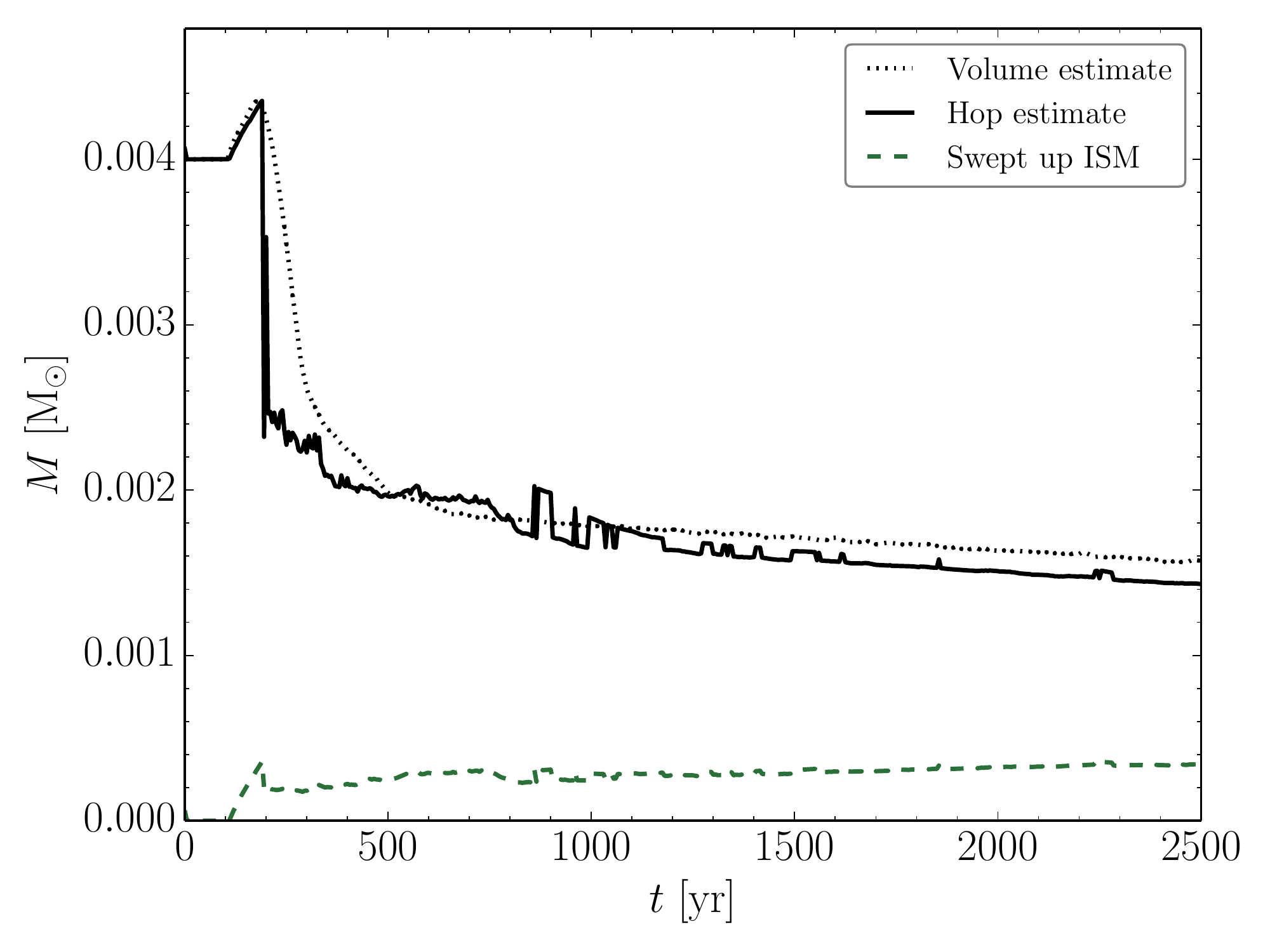}\label{fig:disc_recognition_eda1a}}\hfill
    \subfloat[(b)]{\includegraphics[width=0.49\textwidth]{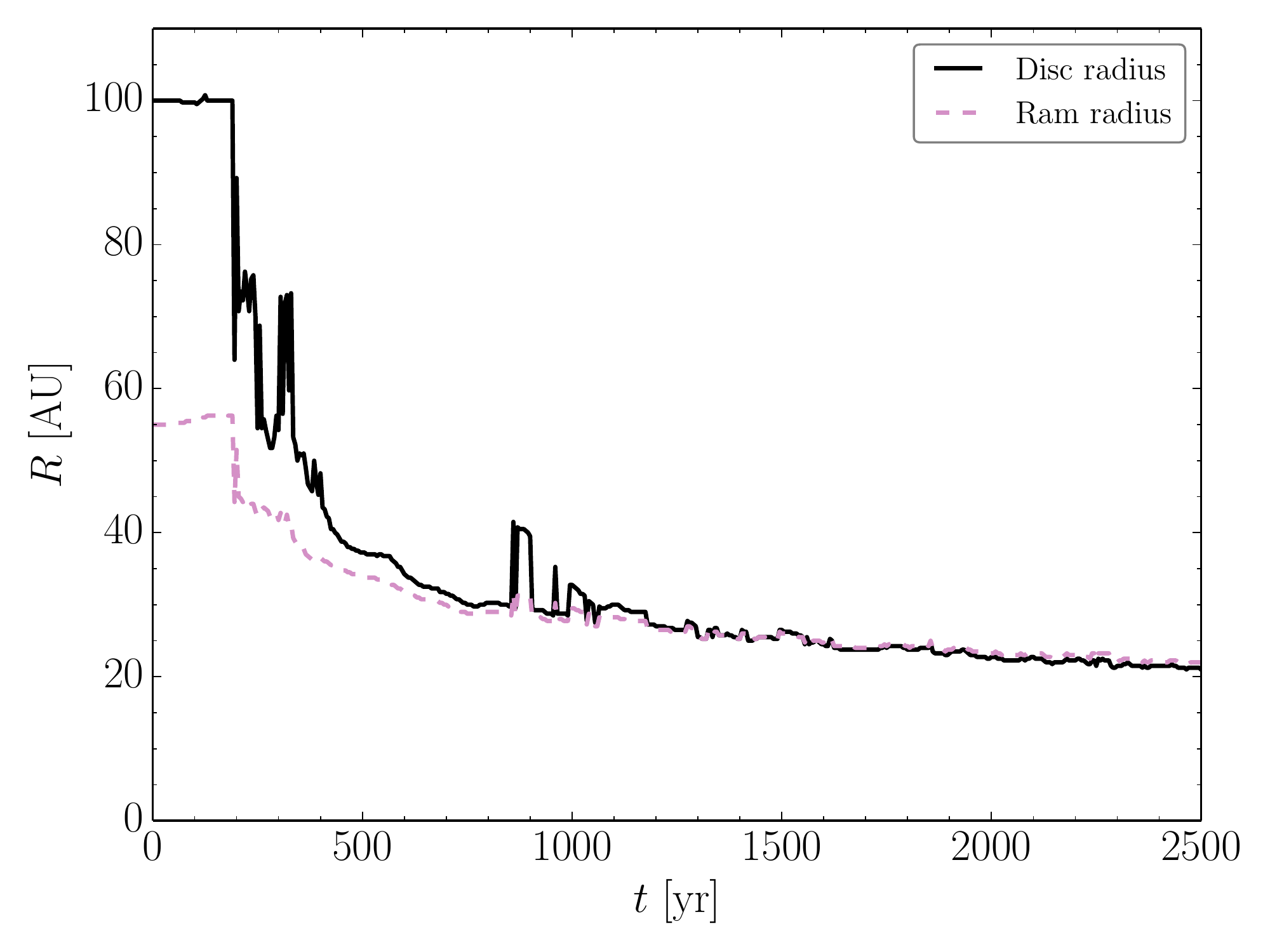}\label{fig:disc_recognition_eda1b}}\\
    \subfloat[(c)]{\includegraphics[width=0.49\textwidth]{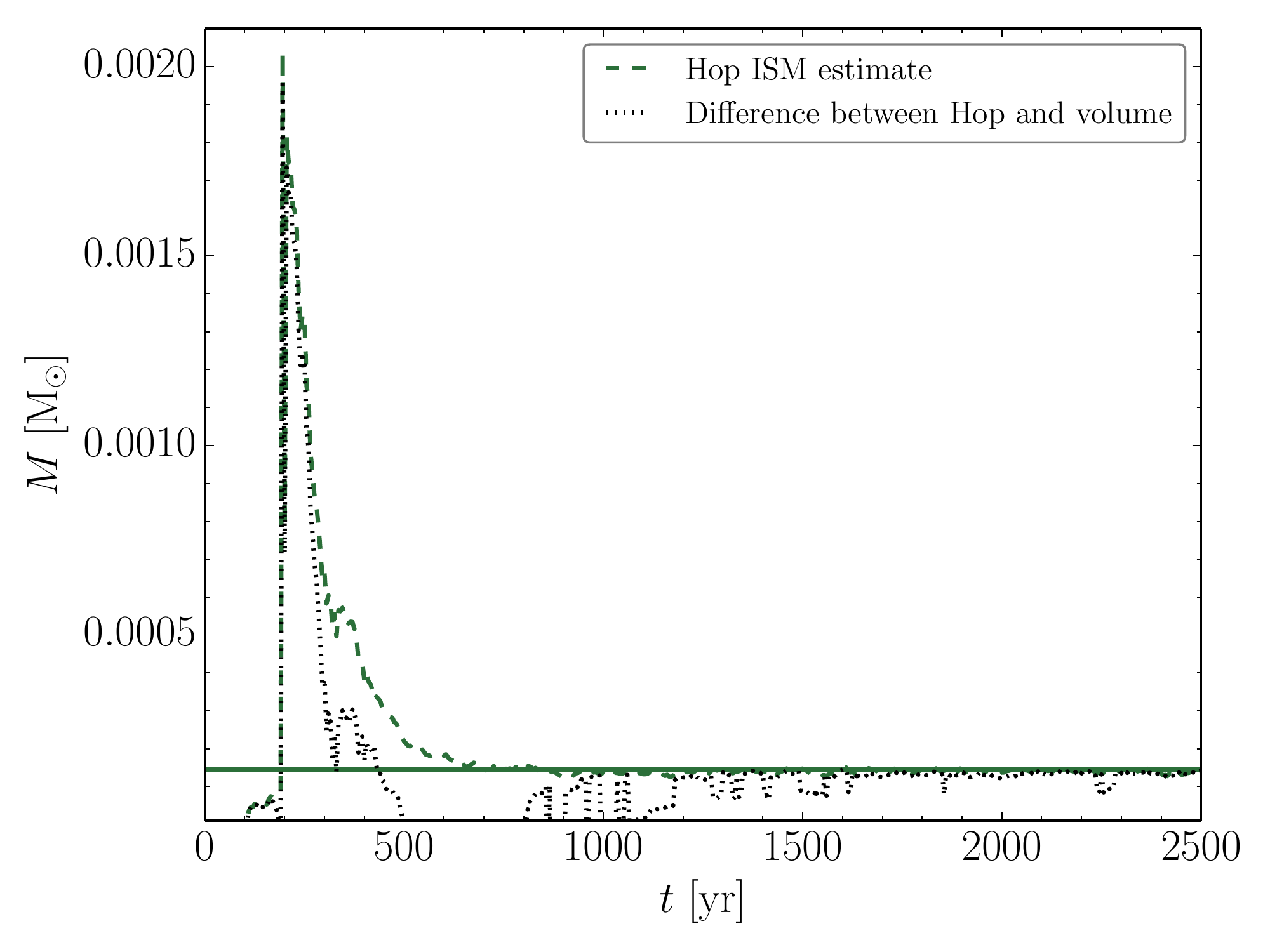}\label{fig:disc_recognition_eda1c}}\hfill\begin{minipage}[b]{0.425\textwidth}\caption{\textbf{\emph{(a):}} The mass of the disc determined in 2 different ways: (1) By drawing a cylinder around the disc and adding the mass of all the particles in that volume (dotted black line) and (2) from the Hop algorithm we use (solid black line, see Sec. \ref{sec:disc_recognition_eda1}). The green dashed line shows the cumulative swept-up ISM mass, i.e.the sum of ISM in the disc and ISM that has been accreted onto the star, as derived with the Hop algorithm. \textbf{\emph{(b):}} The radius of the disc (solid black line) and the radius beyond which the ram pressure dominates, both derived from the surface density profile (see Sec. \ref{sec:disc_recognition_eda1}). \textbf{\emph{(c):}} The solid green horizontal line gives the mass of the volume around the disc assuming it is completely filled with ISM. The dashed green line gives the mass of ISM that is in the volume but not in the disc, as derived with the Hop algorithm. The dotted black line gives the mass of ISM in the volume by calculating the difference between the total mass in the volume and the hop estimate for the disc, i.e. the difference between the solid black line and the dotted black line from the top figure. After $\sim$ 1500 years, the methods agree very well. \label{fig:disc_recognition_eda1}}\end{minipage}\\%
\end{figure*}
To determine how much ISM has been swept-up by the disc at any moment in time, we need to differentiate between the disc and the ISM flow. This is not straightforward since there is no clear separation between the outer edge of the disc and the flow. In order to distinguish between the disc and the ISM flow, we use a clump finding algorithm to identify different groups in the parameter space of the angular speed ($v_{\theta}$), the density ($\rho$) and the speed along the axis of the cylinder($v_x$) of each particle. In this parameter space, we expect the ISM particles to cluster around values of, respectively, 0 km/s, $\rho_{\rm ISM}$ and $v_{\rm ISM}$, while the disc particles will not. For the clump finding algorithm we use Hop \citep{eisenstein98}.

Fig. \ref{fig:disc_recognition_eda1a} shows a comparison of how well this algorithm (solid black line) performs for our reference model compared to drawing a coaxial cylinder around the disc, with a length of 100 AU and a radius of 120 AU, and adding the mass of all the SPH-particles in that volume (dotted black line). The difference between these two lines can be attributed almost completely to the ISM that is in the volume, but not part of the disc. To demonstrate this, Fig. \ref{fig:disc_recognition_eda1c} shows the difference between these two methods (dotted black line), the mass of ISM in the volume as determined with the Hop algorithm (green dashed line) and assuming the volume is completely filled with ISM (solid green horizontal line). In the beginning of the simulation, when the disc is stripped of its outer parts by the ram pressure exerted by the flow, the algorithm has some difficulties differentiating the disc from the ISM flow, but as soon as a stable situation is reached and the stripped outer edges of the disc have left the computational domain, it performs very well. The few spikes visible in Fig. \ref{fig:disc_recognition_eda1} are artefacts: some particles are marked as part of the disc, while they are either being stripped from the outer edge (and already carry some angular momentum) or flow along the outer edge of the disc (and pick up some angular momentum from the disc). These artefacts do not influence our results as they are smoothed out when we linearly fit the data from the moment a steady state is reached.

Once we have determined which particles belong to the disc, we can also determine the surface density profile, $\Sigma(r)$, and radius, $R_{\rm disc}$, of the disc. In order to do so, we calculate the column density of the disc, viewing the disc face-on. We then bin the obtained 2-D surface density in concentric annuli, which gives us the surface density profile as a function of the radius. At $t=0$, we determine $\Sigma(100 {\rm AU})$, i.e. at the initial outer radius of the disc. At consecutive times, we recalculate the surface density profile and consider the radius at which $\Sigma(r)\big|_{t} =\Sigma(100 {\rm AU})\big|_{t=0}$ to be the radius of the disc at that moment in time. The snapshots of our reference model, shown in Fig. \ref{fig:snapshots_eda1}, illustrate that this method performs very well if the simulation has reached a steady state. Furthermore, the radius determined in this way also agrees with the estimate of the truncation radius due to ram pressure stripping, Eq. \ref{eq:ramradius_eda1}, as can be seen in Fig. \ref{fig:disc_recognition_eda1b} where we have plotted both radii as a function of time. Much in the same way as we did for the disc radius, the truncation radius in Fig. \ref{fig:disc_recognition_eda1b} is derived from the surface density profile at each moment in time.

\subsection{Angular momentum conservation}
\label{sec:am_conservation_eda1}
For TreeSPH codes like Fi and Gadget2 angular momentum is not conserved exactly. The reasons for this are that 1) the gravity forces for a Barnes-Hut tree code are not exactly symmetric for particles in different parts of the domain and 2) interactions between particles in different levels of the time step hierarchy are not exactly symmetric. To make sure that the loss of angular moment that we measure is not due to (the lack of) angular momentum conservation, we determined how well angular momentum is conserved in our reference model. In order to do so, we look at the change of total angular momentum, i.e. of all particles that have been in the computational domain over the course of the simulation, between 1500 and 2500 years. The total angular momentum fluctuates around a mean value, without any increasing or decreasing trend with time. To establish a measure of angular momentum conservation, we determined the mean total angular momentum during the last 1000 years of the simulation, which is $1.5\times10^{44}$\,kg\,m$^2$\,s$^{-1}$, while the standard deviation is $1\times10^{40}$\,kg\,m$^2$\,s$^{-1}$. Thus angular momentum is conserved up to about 1 part in $10^4$. We compare this to the angular momentum loss of the disc in Sec. \ref{sec:reference_eda1}.

\section{Results}
 
\begin{table}
\centering
\caption{Different models for the convergence and consistency study}
\label{tb:convergence_eda1}
\begin{tabular}{llccc}
\hline
$\mathbf{N_{\rm disc}}$&\multicolumn{2}{c}{\textbf{SPH Code}}\\
\cline{2-5}
($\mathbf{10^3}$)&\textbf{Gadget2}&$\mathbf{N_{\rm sim}}$&\textbf{Fi}&$\mathbf{N_{\rm sim}}$\\
\hline
4&G4&4&F4&1\\
8&G8&2&F8 &1\\
16&G16&1&F16& 1\\
&G16NF (no ISM inflow)&1&-&\\
&G16CV (constant $\alpha_{\rm SPH}$)&1&-&\\
32&G32&1&-&\\
64&G64&1&-&\\
128&G128&1&-&\\
\hline
\end{tabular}\\
{\em Columns 1 to 5}:  The number of disc particles in a simulation, the label of that simulation and the number of runs of that simulation, for Gadget2 and Fi respectively. Non-standard assumptions are mentioned between brackets. For every run we use the basic set-up as discussed in Sec. \ref{sec:set-up_eda1}. Note that the total number of particles in each simulation is a factor of 7 higher.
\end{table}
We have performed a number of simulations with both Gadget2 and Fi at different resolutions, as listed in Table \ref{tb:convergence_eda1}. The main difference between both codes is the treatment of the artificial viscosity (see Sec. \ref{sec:visc_eda1}). Our reference model is a simulation with Gadget2 and 16.000 disc particles, labelled G16. We discuss our reference model in Sec. \ref{sec:reference_eda1} and we use the other models for the convergence and consistency study which we discuss in Sec. \ref{sec:convandcons_eda1}. Whenever we halve the number of disc particles, we double the number of simulations. We compare these results with runs from the SPH code Fi, which we have performed once for each resolution\footnote{We have done only one simulation for each resolution with Fi, because it is computationally more expensive. 16000 disc particles is the highest resolution we could simulate with Fi in a reasonable amount of time.}. We finish the comparison between both SPH codes by discussing a run of Gadget2 with the same artificial viscosity implementation as used in Fi.

We choose simulation G16 as our reference model, so that we can compare our results with those of M09, who used the same code. We have not chosen one of the Gadget2 simulations with a higher resolution as our reference model, because we can not compare this model to a simulation with the same number of disc particles with Fi. As a benchmark for the accretion rate onto the star and the mass loss from the outer edge of the disc, we have performed a simulation of our reference model without inflow of ISM, labelled G16NF. We discuss this simulation first.

\subsection{Reference model without inflow of ISM}
\label{sec:noflow_eda1}
We have performed a simulation, labelled G16NF, of a protoplanetary disc without inflow of ISM to gauge the mass loss rates at the inner and outer edge of the disc. We compare the mass and angular momentum loss of subsequent simulations to the quantities derived for this simulation. The results of this simulation are summarized in Table \ref{tb:gadget_eda1}, which contains the summary of the simulations with Gadget2. The mass and angular momentum change rates are determined by plotting the concerning quantity as a function of time, e.g. the mass of the disc, and fitting a linear function to it from the moment the simulation reaches a steady state, i.e. from 1500 years onward. Every rate has been defined in this way, i.e. by a linear fit between 1500 and 2500 years. We find an accretion rate onto the star, $\dot{M}_{\rm star}$, of $4.5\times10^{-8}\,\Msun/$yr. 
This is similar to what is observed for stars of $0.4\,\Msun$ \citep{muzerolle05} and is theoretically expected for a disc in a steady state and constant $\alpha_{\nu}$, corresponding to $\alpha_{\rm SPH}=0.1$ \citep{shakura73, armitage11}. We note that M09 found an accretion rate onto the star of $1.5\times10^{-7}\,\Msun/$yr in their isolated disc simulation. 

Roughly half of the mass loss of the disc is due to accretion onto the star, because the inner edge of the disc migrates inwards. This can be seen in Fig. \ref{fig:surface_density_eda1c}, which shows the azimuthally averaged surface density at three different moments in time, i.e. at $t=0$ (solid black), 1500 (dashed purple) and 2500 years (dotted green). The other half is `lost' due to outward spreading of the outer edge as it leaves our computational domain, i.e. when the radial distance of the SPH-particle to the star is more than 500 AU. The outward spreading can be inferred from the decreasing surface density profile at $R>60\,$AU in Fig. \ref{fig:surface_density_eda1c}. The rates and time scales that are quoted  for model G16NF in Table \ref{tb:gadget_eda1} under stripping and angular momentum loss are therefore actually associated with viscous spreading of the disc. We have determined the time scales for viscous spreading by taking the total disc mass and angular momentum at the beginning of the interval and dividing it by the $\dot{M}_{\rm disc}$ and $\dot{J}_{\rm disc}$ respectively. According to this extrapolation, the disc would viscously dissipate on a time scale of $7\times10^4$ years. This is almost half of the numerical viscous time scale at 100 AU, see Sec. \ref{sec:visc_eda1}, but should be considered to be more representative for the disc as a whole.



\subsection{Reference model}
\label{sec:reference_eda1}
\begin{figure*}
\centering
\includegraphics[width=\textwidth]{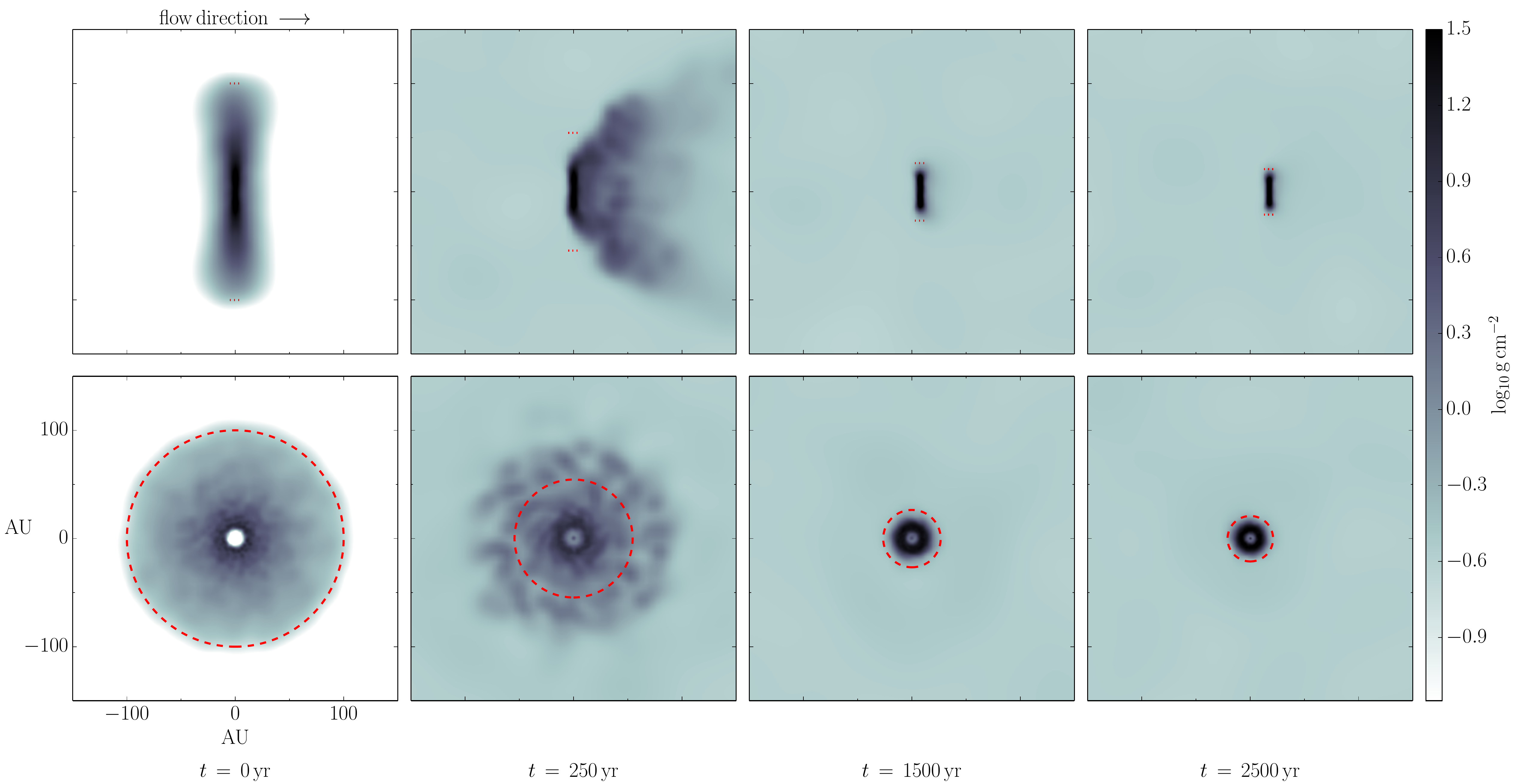}\label{fig:snapshots_eda1a}
\caption{Edge-on (top row) and face-on (bottom row) snapshots from the simulations with our reference model. The column density is shown in logarithmic scale and integrated along the full computational domain. The spatial scale is indicated in the bottom left and the flow direction on the top left. The red circle (bottom row) and dotted bars (top row) indicate the size of the disc as determined from its radius (see Sec. \ref{sec:disc_recognition_eda1}).}\label{fig:snapshots_eda1}
\end{figure*}
Fig. \ref{fig:snapshots_eda1} shows snapshots from the simulations with our reference model. At $t=250$ yr, we can see the ram pressure at play as the flow drags along disc material from radii larger than $R_{\rm trunc}$, which is 55 AU in this case. The third column shows that the simulation has reached a steady state at $t=1500$ yr in which the disc is continuously accreting. The last snapshot illustrates that the disc is shrinking in size during the steady state.
In Fig. \ref{fig:disc_recognition_eda1a} we have plotted the mass of the disc in simulation G16 (solid black line) and the total amount of swept-up ISM (green dashed line) as a function of time. The swept-up ISM is defined as the sum of the ISM that is in the disc at each moment in time and the ISM that has been accreted by the star up to that moment. The total amount of swept-up ISM increases and we can determine the ISM loading rate from the slope of this line, as described in Sec. \ref{sec:noflow_eda1}. This gives a value of $\dot{M}_{\rm ISM} = 1.03\times10^{-7}\,\Msun/$yr. We can compare this value to the rate that we expect based on a simple geometric estimate:
\begin{equation}\label{eq:mdot_eda1}
 \dot{M}_{\rm ISM}=\rho_{\rm ISM} v_{\rm ISM} \pi R_{\rm disc}^2
\end{equation}
where $\rho_{\rm ISM}$ and $v_{\rm ISM}$ are the density and velocity of the ISM respectively and $R_{\rm disc}$ the radius of the disc. The ISM loading rate in the simulation is a factor of about two lower than the geometric rate. We will show in Sec. \ref{sec:convergence_eda1} that the difference is independent of the resolution of our simulation or the code we used. 

\begin{figure*}
        \centering
        \subfloat[(a)]{\includegraphics[width=0.49\textwidth]{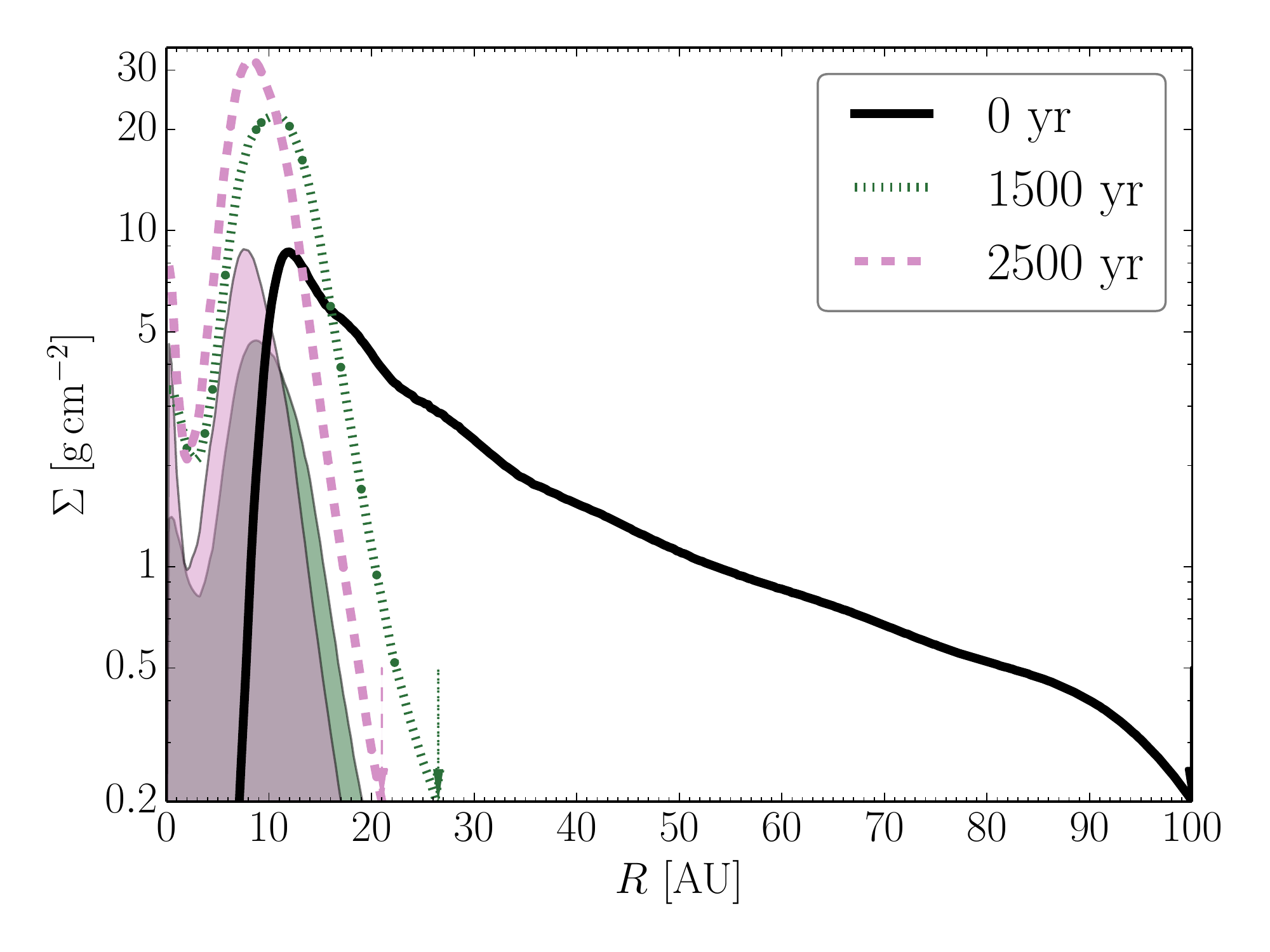}\label{fig:surface_density_eda1a}}\hfill
        \subfloat[(b)]{\includegraphics[width=0.49\textwidth]{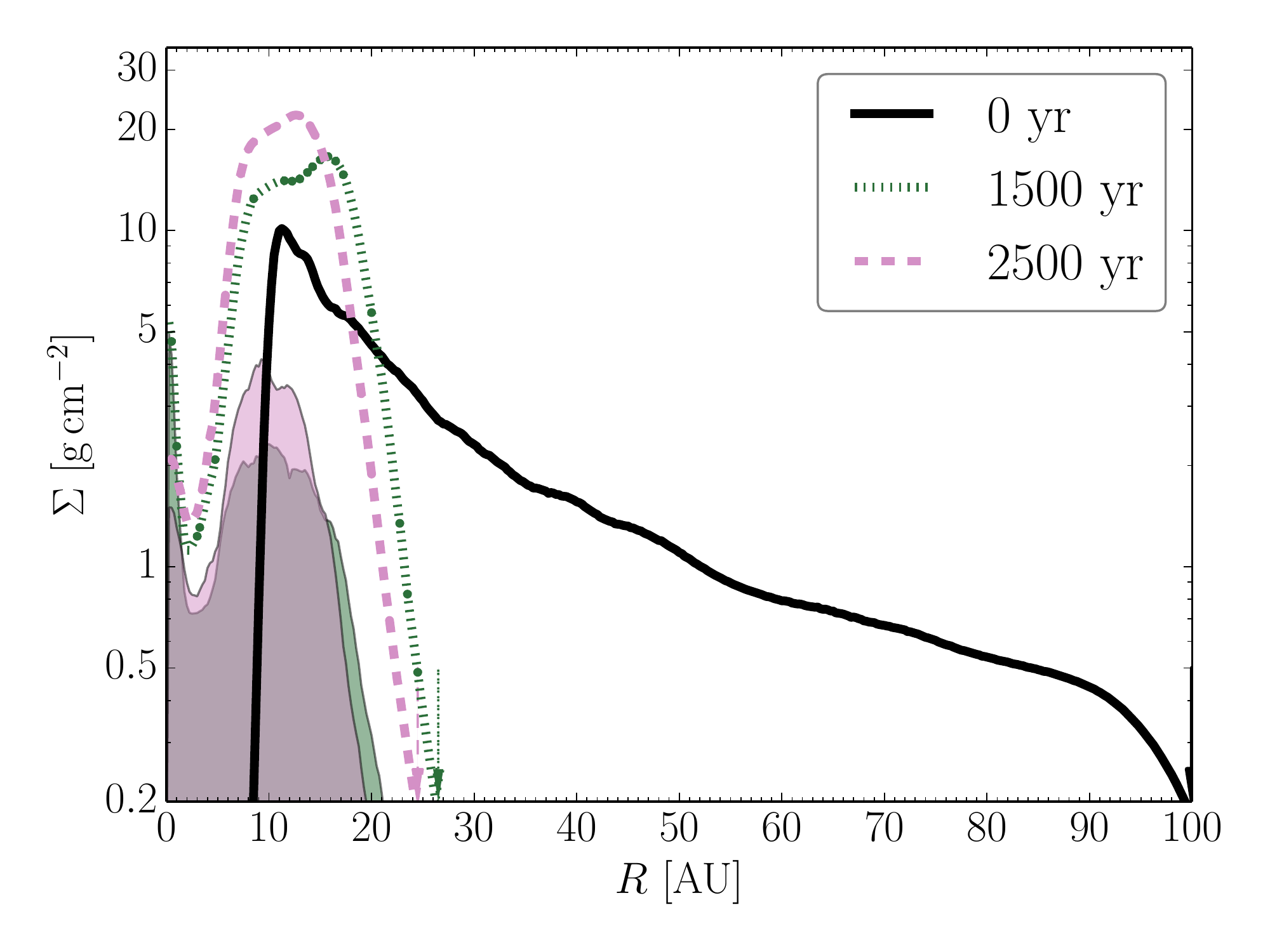}\label{fig:surface_density_eda1b}}\\
        \subfloat[(c)]{\includegraphics[width=0.49\textwidth]{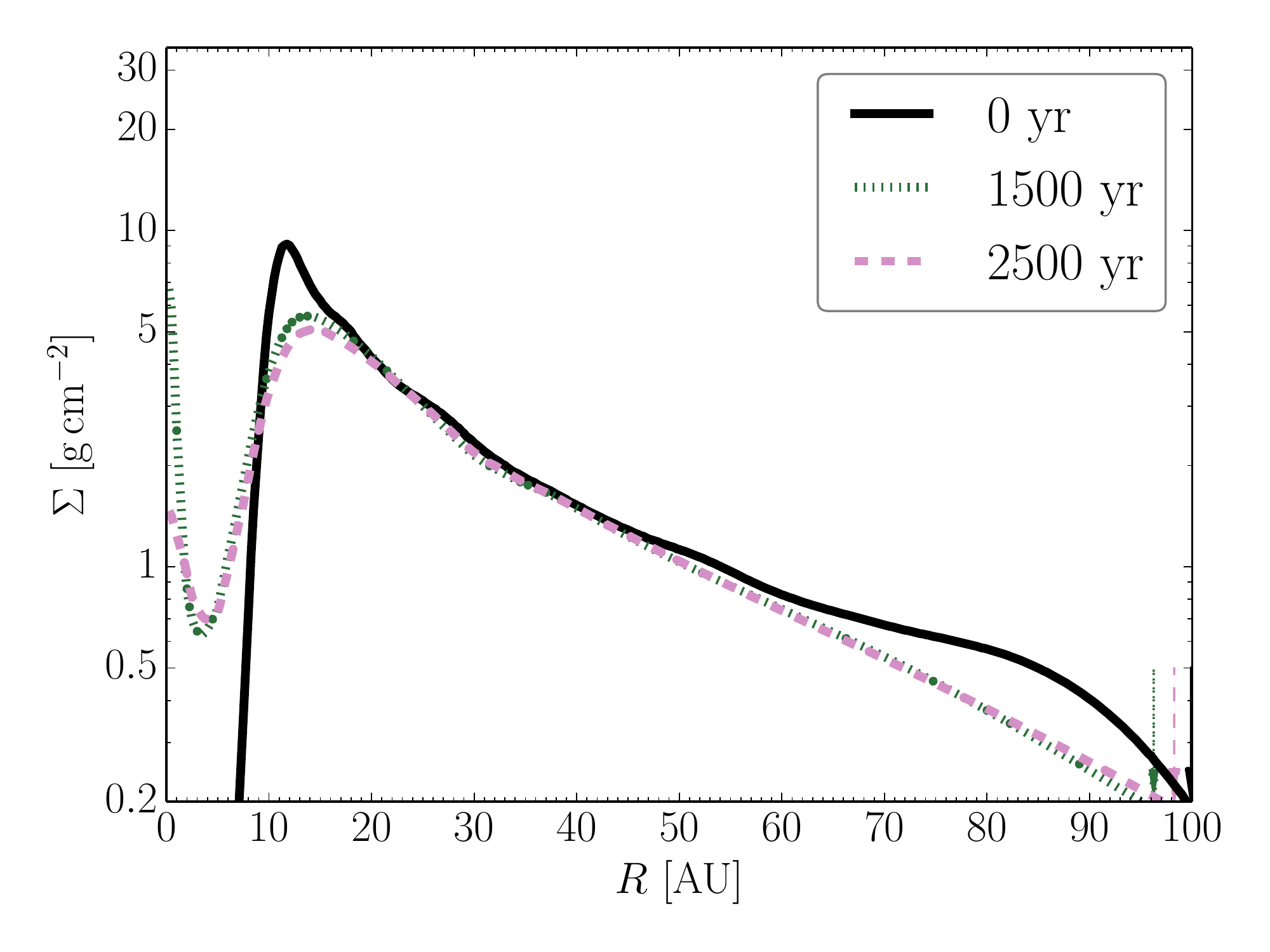}\label{fig:surface_density_eda1c}}\hfill\begin{minipage}[b]{0.425\textwidth}
    \caption{\textbf{\emph{(a):}} The azimuthally averaged surface density profile for the G16 simulation, $\Sigma(r)$, at the start (solid black line), after 1500 years (dashed purple line) and after 2500 years (dotted green line). The contribution of swept-up ISM at each moment is plotted in the corresponding transparent colour. The vertical arrows in the same colour as the surface density profile indicate the radius of the disc at that moment. \textbf{\emph{(b):}} Same as the left figure, but for the F16 simulation. The initial inner radius at 10 AU is better resolved by Fi than by Gadget2. \textbf{\emph{(c):} Same as figure (a), but for the reference model without inflow of ISM.}\label{fig:surface_density_profiles_eda1}}\vspace{1.5cm}\end{minipage}\\%
\end{figure*}

The result that the ISM loading rate is consistently a factor of two lower than given by Eq. \ref{eq:mdot_eda1}, can be understood because at the outer edge of the disc ISM is not entrained by the disc. When the ISM flow first hits the disc, at $t \approx 100$ years, the steep increase of swept-up ISM does agree with the theoretical rate. This is because initially almost all ISM colliding with the surface of the disc is considered part of the disc, seen as an increase in the mass of the disc in Fig. \ref{fig:disc_recognition_eda1a} at $t\approx 100$ years. As the outer edges of the disc are dragged along with the flow, these regions are no longer associated with the disc and there is a large jump in the disc mass and a corresponding smaller one in the swept-up ISM. From that moment on, the effective cross section of the disc, i.e. the area with which it sweeps up ISM, is smaller than the actual surface area of the disc. This is illustrated in Fig. \ref{fig:surface_density_eda1a}, which shows the evolution of the surface density profile for the G16 simulation. The contribution of ISM to the surface density profile at each moment is indicated with filled regions in the corresponding colour. This illustrates that ISM is only entrained by the inner regions and not by the outer regions of the disc.

Although the disc continuously sweeps up ISM, it actually loses mass, see Fig. \ref{fig:disc_recognition_eda1a}, at a rate of $1.68\times10^{-7}\,\Msun/$yr, which is faster than the rate at which the disc sweeps up ISM. The disc loses mass at the inner edge due to accretion onto the star and at the outer edge due to continuous stripping. The accretion rate onto the star is $1.27\times10^{-7}\,\Msun/$yr, roughly equal to the ISM loading rate and almost three times the accretion rate onto the star for an isolated disc. Sweeping up ISM thus enhances the accretion rate onto the star.
The remaining mass is lost from the outer edges at a rate of $1.44\times10^{-7}\,\Msun/$yr. Not only is the ISM not entrained by the outer edge, it is actually removing mass from those regions. This result can be explained in the following way. Due to viscous torques within the disc, as discussed in Sec. \ref{sec:am_redistribution_eda1}, disc material migrates inwards. In order to conserve the total angular momentum of the disc, some disc material moves outwards. This outward diffusion lowers the surface density profile at the outer edge of the disc and material is therefore continuously stripped from the disc by the ram pressure of the ISM flow. The rate at which material is lost from the outer edge of the disc is four times higher than in the isolated case. The rate of angular momentum transport in the disc depends on the (artificial) viscosity and will be discussed in more detail in section \ref{sec:consistency_eda1}, where we compare our two different methods of modelling the viscosity. 

The ablation at the outer edge of the disc causes both mass and angular momentum to be lost from the disc. During the last 1000 years of the simulation, the angular momentum lost by the disc equals $5.43\times10^{42}$\,kg\,m$^2$\,s$^{-1}$, which is more than 460 $\sigma$, as determined in Sec. \ref{sec:am_conservation_eda1}, and thus highly significant in comparison with errors in the angular momentum conservation in the code. Therefore, the loss of angular momentum of the disc can not be ascribed to errors in time integration, but must be due to (the modelling of) physical processes in our simulation. The angular momentum lost due to accretion onto the star in the same time interval, $1.4\times10^{40}$\,kg\,m$^2$\,s$^{-1}$, is insignificant compared to the loss of angular momentum from the outer edge of the disc. In Sec. \ref{sec:convergence_eda1} we discuss that at least half of the angular momentum lost from the disc is caused by interaction with the ISM at the outer edge of the disc in which angular momentum is transferred to the ISM and carried away in the flow. The remaining angular momentum loss is due to continuous stripping of original disc material from the outer edge of the disc. The angular momentum loss suggests a disc lifetime of $6\times10^3$ years, which is an order of magnitude shorter than the time scale derived in the simulation without flow.

\subsubsection{Evolution of the surface density profile}
\label{sec:sdp_eda1}
The slope of the surface density profile at a given radius in Fig. \ref{fig:surface_density_eda1a} increases with time as more ISM is entrained and disc material is transported inwards due to viscous evolution. From the start of the simulation, the artificial viscosity in the mid-plane of the disc increases and is higher than the initial value of 0.1. During the simulation, the typical value of $\alpha_{\rm SPH}$ in the mid-plane of the disc is 0.6. This means that the transport of mass and angular momentum through the disc in the simulation is faster than estimated in Sec. \ref{sec:visc_eda1}. In Sec. \ref{sec:consistency_eda1} we discuss how this compares to a simulation in which $\alpha_{\rm SPH}$ remains equal to 0.1 in the disc.

The purple and green filled regions show the contribution of ISM in the disc to the surface density profile at $t=1500$ and 2500 years. At each snapshot, the relative contribution of the ISM to the surface density profile is highest at small radii and decreases towards larger radii. The ISM in the disc migrates inwards due to viscous torques and thus follows the same trend as the total surface density profile of the disc. Furthermore, new ISM with no angular momentum is continuously entrained by the disc, which also contributes to the inward migration of disc material. Both processes, continuous accretion and viscous evolution, steepen the exterior slope of both the total and ISM surface density profile. The total fraction of ISM in the disc increases with time as more material is swept-up.

\subsection{Convergence and consistency}
\label{sec:convandcons_eda1}
\begin{figure*}
        \centering
        \subfloat[(a)]{\includegraphics[width=0.49\textwidth]{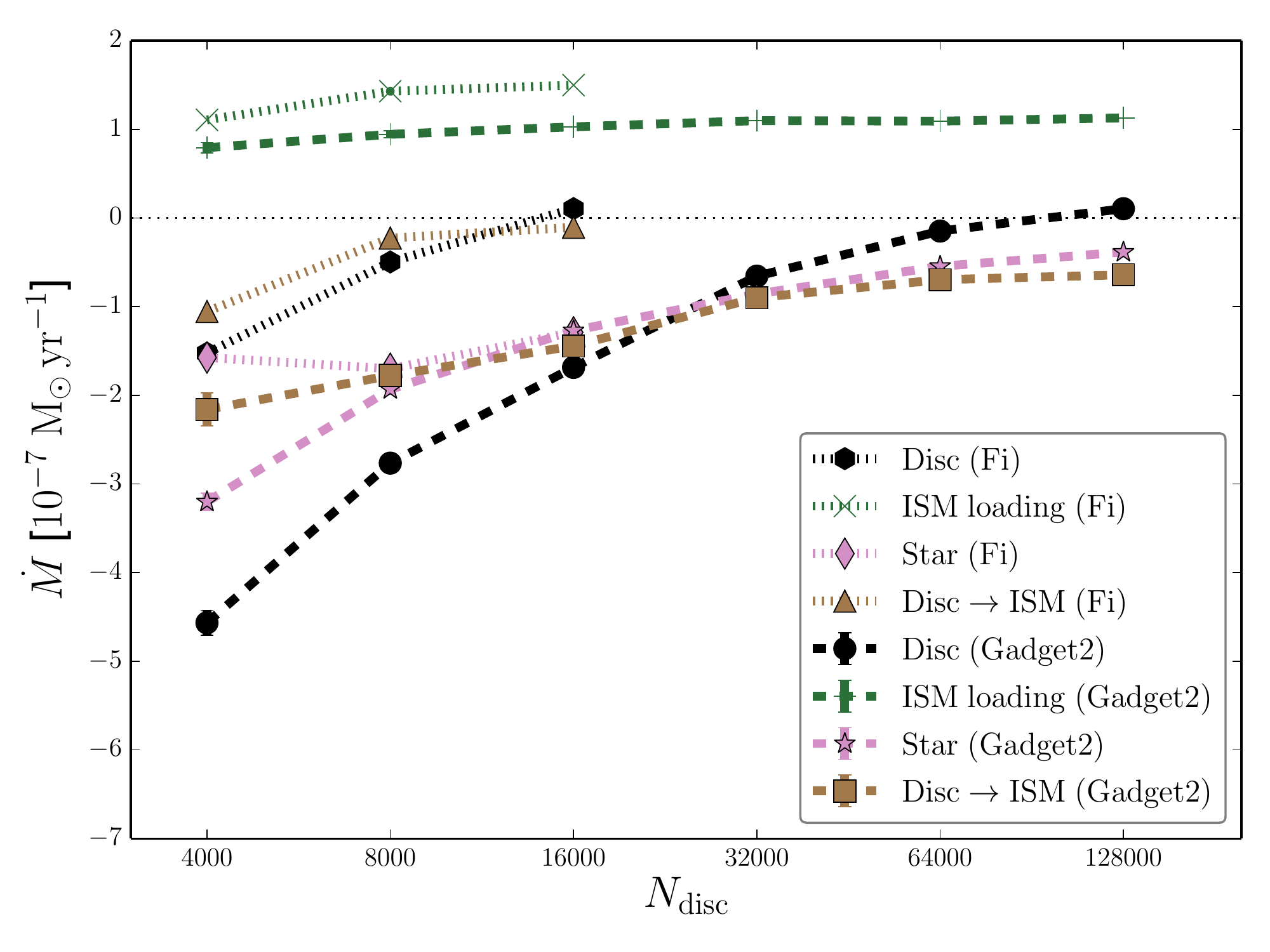}\label{fig:convergence_rates_eda1a}}\hfill
        \subfloat[(b)]{\includegraphics[width=0.49\textwidth]{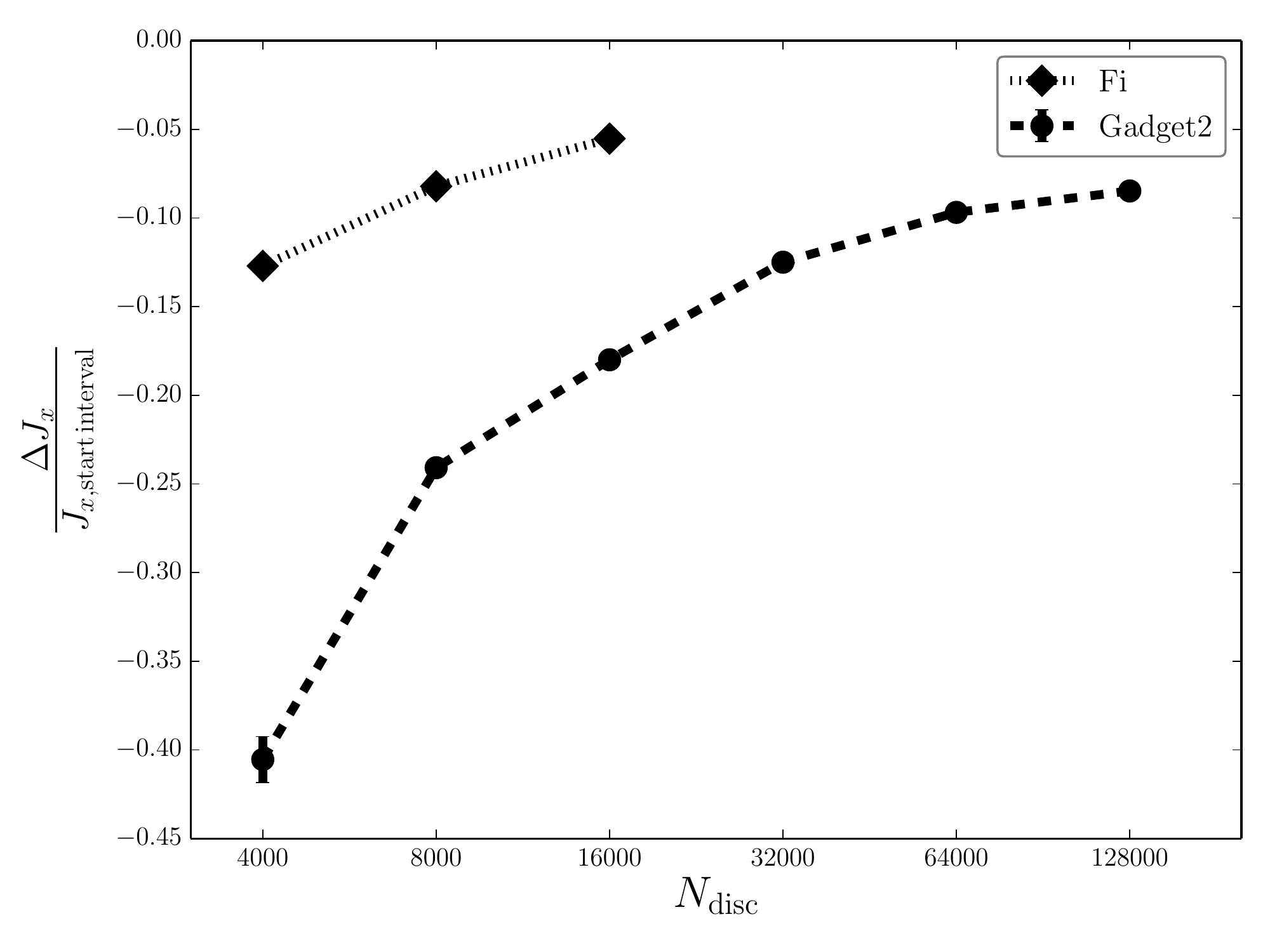}\label{fig:convergence_rates_eda1b}}\\
	\subfloat[(c)]{\includegraphics[width=0.49\textwidth]{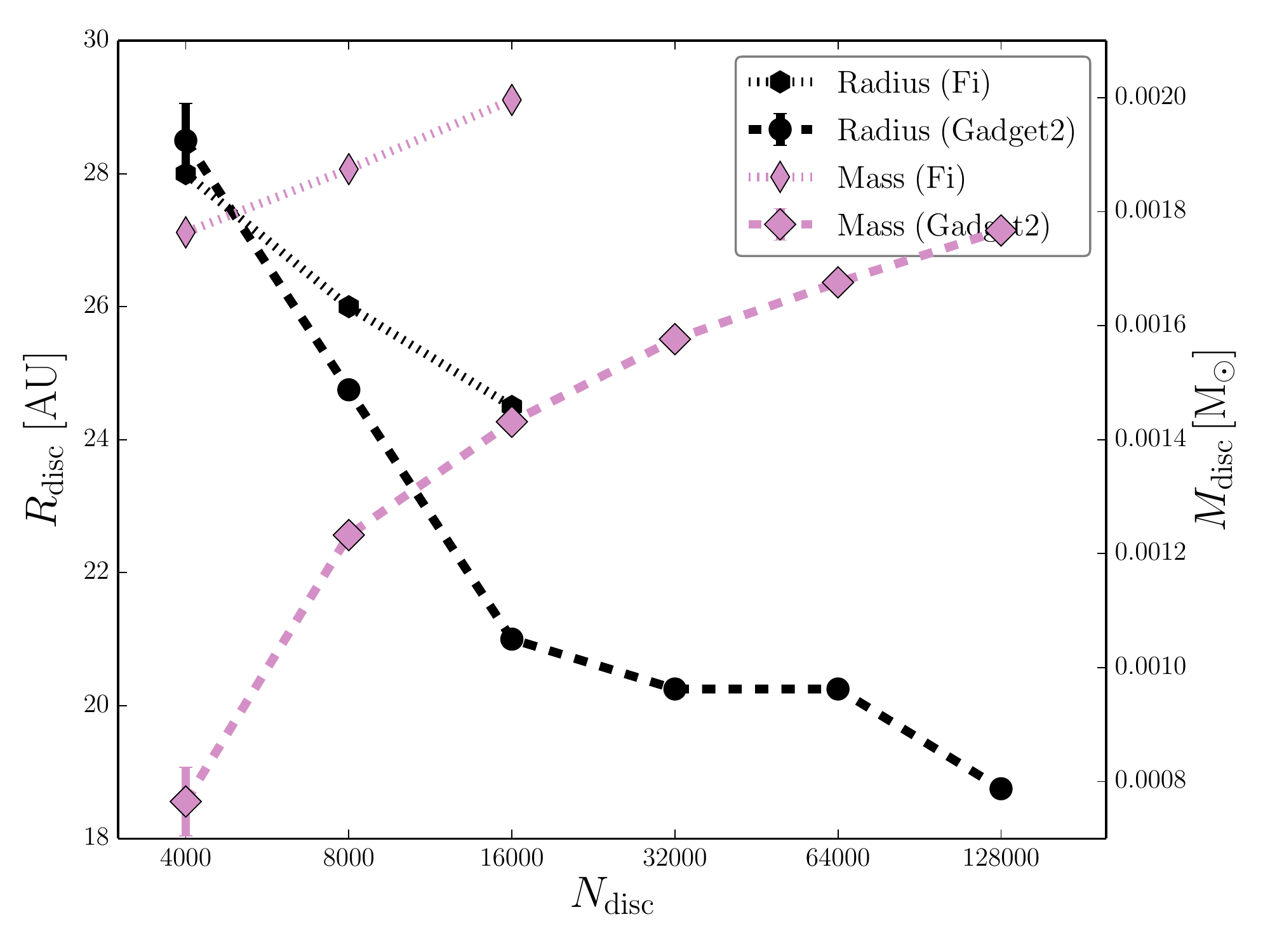}\label{fig:convergence_rates_eda1c}}\hfill\begin{minipage}[b]{0.425\textwidth}\caption{\textbf{\emph{(a):}} Average mass loss and gain rates, $\dot{M}$, as a function of the number of disc particles, N$_{\rm disc}$ for both SPH-codes Fi (dotted) and Gadget2 (dashed).
The change of the disc mass (black) consists of three components: mass is lost due to accretion onto the sink particle (purple) and stripping at the outer edge (brown), and gained from sweeping up ISM by the disc and star (green). The rates are determined by a linear fit (see Sec. \ref{sec:convergence_eda1}). \textbf{\emph{(b):}} The loss of angular momentum of the disc during the final 1000 years interval as a fraction of its total angular momentum plotted against resolution for Fi (dotted) and Gadget2 (dashed). We only consider the angular momentum component along the (initial) rotation axis of the disc. \textbf{\emph{(c):}} The radius (black) and mass (purple) of the disc at the end of the simulation as a function of the number of disc particles, N$_{\rm disc}$ for both SPH-codes Fi (dotted) and Gadget2 (dashed). The determination of these quantities is described in Sec. \ref{sec:disc_recognition_eda1}. \label{fig:convergence_rates_eda1}}\end{minipage}\\%
\end{figure*}

In this section we compare the outcome of simulations with different numbers of disc particles and SPH codes. 
The runs for the convergence study are listed in Table \ref{tb:convergence_eda1}. Fig. \ref{fig:convergence_rates_eda1a} shows the average mass gain and loss rates that are relevant for our study: the accretion rate onto the star, the net rate at which ISM material is swept-up, the rate at which initial disc material is stripped and the total rate of change in the disc mass. The rates are determined by considering the various mass quantities, e.g. the mass of the disc and the star, as a function of time during the last 1000 years of the simulation and applying a linear fit to their slope. We summarize all the quantities plotted in Fig. \ref{fig:convergence_rates_eda1} in Tables \ref{tb:gadget_eda1} and \ref{tb:fi_eda1}. We first discuss the convergence of the simulations, i.e. the effect of changing the number of disc particles, and subsequently the consistency between Gadget2 and Fi. 

\subsubsection{Convergence}
\label{sec:convergence_eda1}

Fig. \ref{fig:convergence_rates_eda1a} shows that, except for the ISM loading rate, all rates have a decreasing trend with increasing resolution, with the exception of the accretion rate onto the star in the low resolution F4 simulation. The decreasing trends of the accretion rate onto the star, the stripping rate and the mass loss of the disc can be understood as follows. When the number of disc particles increases, the average smoothing length of the SPH particles decreases. Therefore, at lower resolution, the viscous time scale is shorter (according to Eq. \ref{eq:reynolds_eda1}) and the transport of mass and angular momentum through the disc is faster than at higher resolutions. Moreover, as we will discuss in Sec. \ref{sec:consistency_eda1}, in our simulations with Gadget2 the artificial viscosity in the disc is higher for lower resolutions. This further increases the dependence of mass and angular momentum transport on the number of disc particles. As discussed in Sec. \ref{sec:reference_eda1}, there is a physical relation between $\dot{M}$ at the inner edge of the disc, i.e. accretion onto the star, and $\dot{M}$ at the outer edge of the disc, i.e. the ablation rate, which is the trend we observe in Fig. \ref{fig:convergence_rates_eda1a}. This implies that the transport of angular momentum in the simulations is dominated by numerical effects, as these rates are sensitive to the number of disc particles and have not converged yet.

The rate of mass change of the disc follows the same trend as the accretion and ablation rate, since the disc loses less mass at both the inner and outer edge with increasing resolution. The ISM loading rate, however, appears to be independent of the number of disc particles and is almost constant at all resolutions. This implies that the effective cross section with which the disc entrains ISM is roughly equal for all resolutions. Fig. \ref{fig:convergence_rates_eda1c} shows that the radius, i.e. total surface area, of the disc is larger for lower resolution, but the outskirts of the disc are also more diffuse in that case. As discussed above, if the outer edge of the disc is too diffuse, the ISM particles are not entrained by the disc. At higher resolution, the disc is more compact, as can be seen from both the decreasing radius and increasing mass in Fig. \ref{fig:convergence_rates_eda1c}. This can again be understood from the concept of mass and angular momentum transport: at high resolution less mass is ablated and more mass remains in the disc, which in turn migrates inwards making the disc more compact. The surface density at the outer edge of the disc is therefore higher for a larger number of disc particles. The net effect across different resolutions is that the effective cross section of the disc remains approximately the same, i.e. the effective cross section depends on the height of the surface density profile in the outskirts of the disc. However, the ratio of the effective cross section over the actual surface area of the disc, $\sigma_{\rm disc}/A_{\rm disc}$, increases with resolution.

\begin{figure}
\centering
    \includegraphics[width=0.49\textwidth]{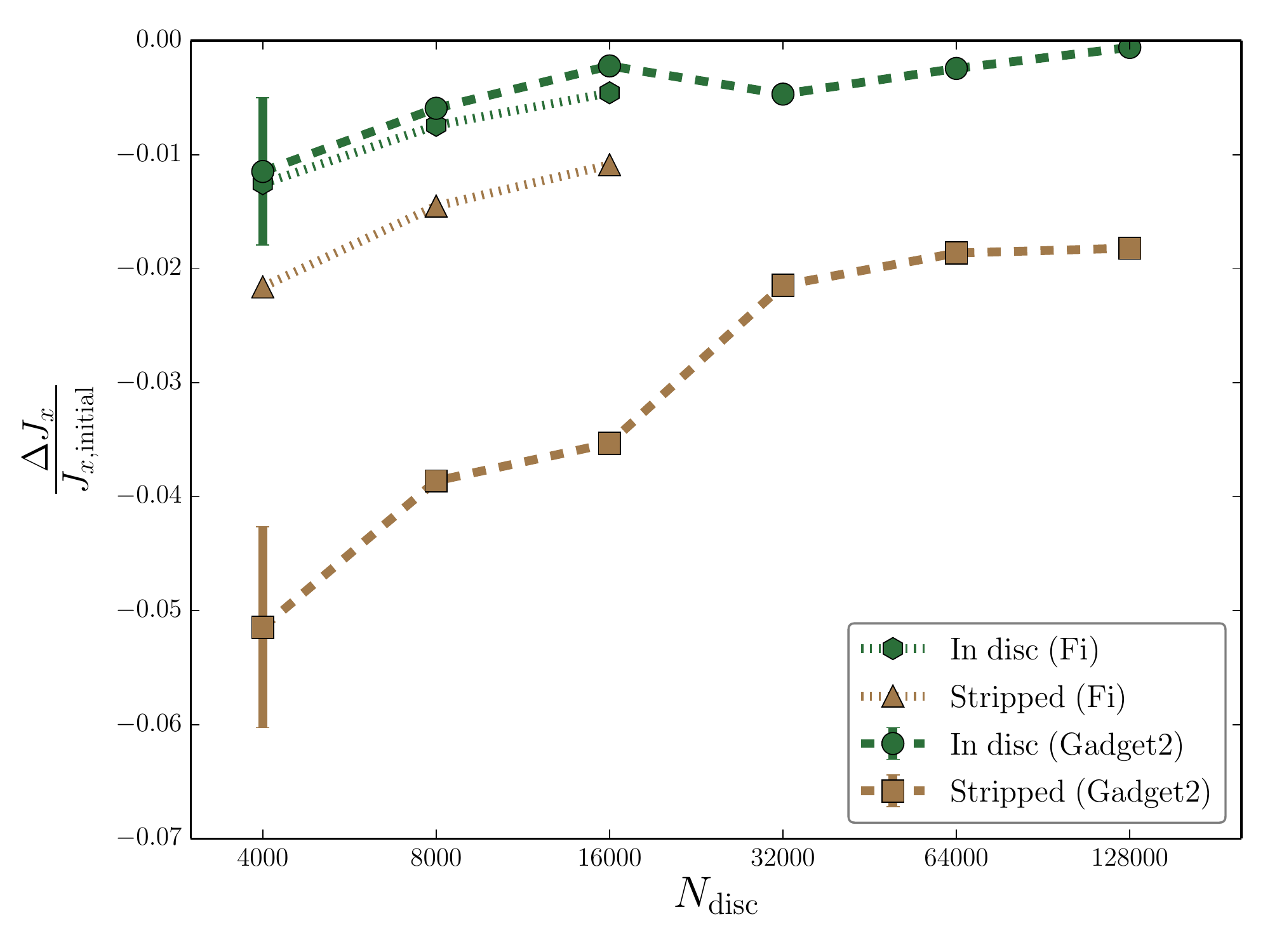}
    \caption{The angular momentum lost from the disc during the final 1000 years in terms of the initial angular momentum of the disc for both Fi (dotted) and Gadget2 (dashed). The angular momentum loss associated with processes in the disc, e.g. accretion of ISM, angular momentum exchange between accreted ISM and original disc material, is plotted in green and angular momentum loss associated with ablation is shown in brown. We have omitted the angular momentum loss due to accretion onto the star, because it is insignificant (as discussed in \ref{sec:reference_eda1}).\label{fig:AMlosscomponents_eda1}}  
\end{figure}

Fig. \ref{fig:convergence_rates_eda1b} shows the amount of angular momentum lost from the disc during the last 1000 years of the simulation as a fraction of its angular momentum at the start of that interval. It shows that more angular momentum is lost at lower resolution. The time scales that can be derived from this angular momentum loss are shown in Tables \ref{tb:gadget_eda1} and \ref{tb:fi_eda1}. These time scales are of the same order as, although mostly consistently smaller than, the time scales derived from the mass loss of the disc.

To better understand how the angular momentum is lost, we have split the angular momentum loss during the final 1000 years into two components in Fig. \ref{fig:AMlosscomponents_eda1}: (1) the angular momentum loss associated with ablation and (2) the angular momentum lost due to processes in the disc, i.e. the accretion of ISM and the exchange of angular momentum between the swept-up ISM and original disc material. The angular momentum loss is normalized to the initial angular momentum of the disc so we can compare the components on an absolute scale. We do not show the angular momentum lost due to accretion onto the star, because it is two orders of magnitude smaller than the angular momentum lost due to ablation (see Sec. \ref{sec:reference_eda1}). Ideally, the second component attributed to processes in the disc should be very small. The swept-up ISM should have no net azimuthal angular momentum and the amount of angular momentum gained by swept-up ISM in the disc should equal the amount that is lost by material that remains in the disc. It is not expected to be exactly zero, as the angular momentum that is lost from the outer edge of the disc has been gained at the expense of material that is still in the disc.

The component related to stripped material can be separated into two constituents: angular momentum carried away by original disc material and angular momentum carried away by ISM that briefly interacts with the disc, gains angular momentum and then moves along with the flow\footnote{The Hop algorithm considers this ISM to have been associated with the disc during a few time steps. This is an artefact of the Hop algorithm, but these interactions do carry away angular momentum from the outer edge of the disc nonetheless. In general it is difficult to truly disentangle all components, also considering that angular momentum stripped from the outer edge of the disc has been gained from material that still resides in the disc. We therefore only provide the total angular momentum loss rates and time scales.}. In our simulations with Gadget2 both constituents contribute equally to the angular momentum loss associated with stripping. However, in our simulations with Fi, 93\,\% of the angular momentum loss associated with stripping is actually due to angular momentum taken away by the ISM. This explains why the angular momentum loss associated with stripping differs by roughly a factor two between both codes. Since stripping dominates the total angular momentum loss of the disc, the same factor of two difference between both codes is also seen in $\dot{J}_{\rm disc}$ in Tables \ref{tb:gadget_eda1} and \ref{tb:fi_eda1} when comparing simulations with the same resolution. We therefore argue that the angular momentum loss of the disc is dominated by frictional interactions between the ISM flow and the outer edge of the disc. The lifetime of the disc in our simulations is thus predominantly limited by angular momentum loss associated with this process. Both codes show a decreasing trend of angular momentum loss with increasing resolution and we will discuss in Sec. \ref{sec:dis_amloss_eda1} to what extent this angular momentum loss is physical. 

\subsubsection{Consistency}
\label{sec:consistency_eda1}
Although the trend of all quantities in Fig. \ref{fig:convergence_rates_eda1} is the same for both codes, those that are determined from the simulations with Gadget2 are consistently lower than, or at most equal to, the same quantity determined from the simulations with Fi, with the exception of the disc radius at the lowest resolution. One of the fundamental differences between Gadget2 and Fi is the treatment of the artificial viscosity: in Fi it is constant, i.e. $\alpha_{\rm SPH}=0.1$, while in Gadget2 it can vary by an order of magnitude depending on the local velocity gradient. As mentioned in Sec. \ref{sec:sdp_eda1}, during the simulations with Gadget2, $\alpha_{\rm SPH}$ is not equal to 0.1 in the disc, as derived from accretion disc theory and observations (see Sec. \ref{sec:visc_eda1}), but approximately 0.6. In fact, $\alpha_{\rm SPH}$ in the mid-plane of the disc increases steeply inwards as a function of the radius between about $\frac{1}{2}R_{\rm disc}$ and the star. At the outer edge of the disc $\alpha_{\rm SPH}$ is also higher, due to the high velocity gradient caused by the shock. The value of $\alpha_{\rm SPH}$ as a function of radius is reflected by the surface density profiles of the G16 simulation in Fig. \ref{fig:surface_density_eda1a}. The radius at which $\alpha_{\rm SPH}$ has a minimum in the G16 simulation corresponds to the peak in the surface density profile. Thus matter accumulates at a radius in the disc where the transport of mass and angular momentum is slowest. In contrast, $\alpha_{\rm SPH}$ is constant in the F16 simulation and the surface density profile is therefore broader and less steep than in the G16 simulation (see Fig. \ref{fig:surface_density_eda1b}). Furthermore, $\alpha_{\rm SPH}$ depends on the number of disc particles: the higher the resolution, the lower $\alpha_{\rm SPH}$ ($\alpha_{\rm SPH} \approx 0.4$ at the highest resolution). Therefore, in a given simulation with Gadget2, more mass is transported inwards, and angular momentum is transported outwards accordingly, than in the same simulation with Fi. Fig. \ref{fig:AMlosscomponents_eda1} shows, as discussed in the previous section, that in the simulations with Fi the angular momentum loss associated with ablation is consistently a factor two lower than in the simulations with Gadget2. This implies that the angular momentum loss not only depends on resolution, but also on the modelling of physical processes. In principle, the smoothing length should be the same in simulations with the same number of particles for both codes. However, since the artificial viscosity is higher at smaller radii in Gadget2, numerical effects start dominating at an earlier stage in the simulation and at these small radii the smoothing length increases as a result of the faster inward movement of particles. At radii smaller than 2 AU, the numerical viscous time scale in the disc becomes smaller than the orbital period. 

Despite these differences, the ISM loading rate in both codes and at all resolutions agree within a factor 1.5, because we are looking at the whole star and disc system. In the case of Fi, accreted ISM resides in the disc for a longer time before it is accreted onto the star. So to determine the ISM loading rate it is less relevant how rapidly material is transported through the disc, as long as it has been swept up by the disc. Both SPH codes agree that the disc will always lose angular momentum, even if the disc gains mass in some simulations (i.e. the F16 and G128 simulations), and that the angular momentum is predominantly lost from the outer edge of the disc. However, the time scales derived from the angular momentum loss of the disc in the simulations with Fi differ by a factor of three from those derived with Gadget2 (see Table \ref{tb:gadget_eda1} and \ref{tb:fi_eda1}). 

\subsubsection{Gadget2 and Fi with the same viscosity implementation}
We have performed a simulation with 16.000 disc particles using Gadget2 with the same (constant) viscosity implementation as in Fi, i.e. $\alpha_{\rm SPH}=0.1$ and $\beta_{\rm SPH}=1$, to address the differences between the results of Gadget2 and Fi. We label this simulation G16CV and summarize the results in Table \ref{tb:gadget_eda1}. The amount of swept-up ISM and the ISM loading rate are almost the same as in the F16 simulation. The main difference between this simulation and the F16 simulation is that both disc and ISM material are accreted faster onto the star in the G16CV simulation than they are in the F16 simulation. As a result the disc is losing mass in the G16CV simulation, even though no original disc material is stripped from the outer edge of the disc, i.e. $\dot{M}_{\rm strip} = 0\,\Msun$/yr. The ISM loading rate in the G16CV simulation is dominated by the accretion rate of ISM onto the star; the amount of ISM in the disc remains roughly constant as a function of time when the system has reached a steady state. This also partly explains the difference with $\dMdtstar$ in the G16 simulation, where the accretion rate onto the star is dominated by the accretion of disc material and therefore a factor two lower. As in the F16 simulation, the disc loses its angular momentum predominantly through interaction with the ISM flow at the outer edge of the disc. However, in the G16CV more angular momentum is transferred from the outer edge of the disc to the ISM flow than in the F16 simulation. Even though the viscosity prescription is the same in both simulations, the inward transport of mass occurs faster in the G16CV simulation. We have not been able to pinpoint the exact cause of the difference between the two codes. It could be that the discrepancies are attributable to other differences in e.g. the implementation of the time-stepping or the limiters on the acceleration, that are more difficult to discern.

\section{Discussion}
\label{sec:dis_eda1}

\begin{table*}
\begin{minipage}{\textwidth}  
\centering
\caption{Quantities of different runs with Gadget2 of which some are plotted in Fig. \ref{fig:convergence_rates_eda1}. \textbf{\emph{Top table from left to right:}} the model, the rate of mass change of the disc, the accretion rate onto the star, the ISM loading rate onto the disc and star system, the rate at which original disc material is stripped from the outer edge of the disc and the estimate for the time scale on which the disc would lose all its mass. \textbf{\emph{Bottom table from left to right:}} the model, the mass of the disc at the end of the simulation ($t=2500$ years), the radius of the disc at $t=2500$ years, the rate at which the disc loses angular momentum and an estimate for the time scale on which the disc would lose all its angular momentum.}
\renewcommand\footnoterule{}
{\begin{tabular}{l|ccccc}
\hline
\textbf{Model}&$\dMdtdisc$ ($\Msun/$yr)&$\dMdtstar$ ($\Msun/$yr)&$\dMdtISM$ ($\Msun/$yr)&$\dMdtstrip$ ($\Msun/$yr)&$\tau_{\dot{M}}$ (yr)\\
\hline
G4&$(-4.6 \pm  0.1)\times10^{-7}$&$(-3.20 \pm  0.1)\times10^{-7}$&$(0.79 \pm  0.06)\times10^{-7}$&$(-2.2 \pm  0.2)\times10^{-7}$&$(1.7 \pm 0.1)\times10^{3}$\\

G8&$(-2.76 \pm 0.07)\times10^{-7}$&$(-1.93 \pm 0.03)\times10^{-7}$&$(0.94\pm 0.04)\times10^{-7}$&$(-1.78 \pm  0.06)\times10^{-7}$&$(4.5 \pm 0.01)\times10^{3}$\\

G16&$-1.68\times10^{-7}$&$-1.27\times10^{-7}$&$ 1.03\times10^{-7}$&$-1.44\times10^{-7}$&$8.5\times10^{3}$\\

G32&$-0.66\times10^{-7}$&$-0.86\times10^{-7}$&$ 1.10\times10^{-7}$&$-0.90\times10^{-7}$&$24.0\times10^{3}$\\

G64&$-0.15\times10^{-7}$&$-0.54\times10^{-7}$&$ 1.09\times10^{-7}$&$-0.70\times10^{-7}$&$114.0\times10^{3}$\\

G128&$0.11\times10^{-7}$&$-0.38\times10^{-7}$&$ 1.13\times10^{-7}$&$-0.64\times10^{-7}$&$\emph{167.5}\times10^{3}$\\
\hline

G16NF\footnote{\label{note1} The stripping rates and time scales derived for this simulation correspond to viscous spreading and not to actual stripping.}&$-0.81\times10^{-7}$&$-0.45\times10^{-7}$&$ -$&$-0.35\times10^{-7}$&$48.0\times10^{3}$\\
\hline
G16CV&$-1.10\times10^{-7}$&$-2.44\times10^{-7}$&$ 1.35\times10^{-7}$&$0\times10^{-7}$&$16.1\times10^{3}$\\
\hline

\multicolumn{5}{c}{ }\\
\hline
&$\Mdisc$ ($\Msun$)&$\Rdisc$ (AU)&$\dot{J}_{\mathrm{disc}}$ (kg\,m$^2$\,s$^{-2}$)&$\tau_{\dot{J}}$ (yr)\\
\hline
G4&$(7.7 \pm 0.6)\times10^{-4}$&$28.5 \pm  0.6$&$(-2.86 \pm  0.09)\times10^{32}$&$(2.5 \pm 0.1)\times10^{3}$\\

G8&$(12.33 \pm 0.02)\times10^{-4}$&$24.8$&$(-2.13 \pm  0.02)\times10^{32}$&$(4.2 \pm 0.1)\times10^{3}$\\

G16&$14.31\times10^{-4}$&$21.0$&$-1.72\times10^{32}$&$5.6\times10^{3}$\\

G32&$15.76\times10^{-4}$&$20.3$&$-1.22\times10^{32}$&$8.0\times10^{3}$\\

G64&$16.76\times10^{-4}$&$20.3$&$-0.98\times10^{32}$&$10.3\times10^{3}$\\

G128&$17.67\times10^{-4}$&$18.8$&$-0.90\times10^{32}$&$11.8\times10^{3}$\\
\hline
G16NF\footnoteref{note1}&$38.61\times10^{-4}$&$98.3$&$-0.67\times10^{32}$&$70.9\times10^{3}$\\
\hline
G16CV&$17.69\times10^{-4}$&$28.75$&$-1.37\times10^{32}$&$9.7\times10^{3}$\\
\hline
\end{tabular}
}
\label{tb:gadget_eda1}

\end{minipage}
\end{table*}

\begin{table*}
\centering
\caption{Same as table \ref{tb:gadget_eda1} but for the simulations with Fi. Note that $\tau_{\dot{M}}$ is italic for the F16 simulation, because the disc actually gains mass in this model.}
{\begin{tabular}{l|ccccc}
\hline
\textbf{Model}&$\dMdtdisc$ ($\Msun/$yr)&$\dMdtstar$ ($\Msun/$yr)&$\dMdtISM$ ($\Msun/$yr)&$\dMdtstrip$ ($\Msun/$yr)&$\tau_{\dot{M}}$ (yr)\\
\hline
F4&$-1.52\times10^{-7}$&$-1.57\times10^{-7}$&$1.10\times10^{-7}$&$-1.06\times10^{-7}$&$11.6\times10^{3}$\\

F8&$-0.50 \times10^{-7}$&$-1.70 \times10^{-7}$&$ 1.43\times10^{-7}$&$-0.23\times10^{-7}$&$37.8\times10^{3}$\\

F16&$0.11\times10^{-7}$&$-1.29 \times10^{-7}$&$1.50\times10^{-7}$&$-0.10\times10^{-7}$&$\emph{187.4}\times10^{3}$\\
\hline

\multicolumn{5}{c}{ }\\
\hline
&$\Mdisc$ ($\Msun$)&$\Rdisc$ (AU)&$\dot{J}_{\mathrm{disc}}$ (kg\,m$^2$\,s$^{-2}$)&$\tau_{\dot{J}}$ (yr)\\
\hline
F4&$17.6 \times10^{-4}$&$28.0$&$-1.60\times10^{32}$&$7.9\times10^{3}$\\

F8&$18.7 \times10^{-4}$&$26.0$&$-1.04\times10^{32}$&$12.2\times10^{3}$\\

F16&$20.0 \times10^{-4}$&$24.5$&$-0.73\times10^{32}$&$18.1\times10^{3}$\\
\hline

\end{tabular}
}
\label{tb:fi_eda1}
\end{table*}
We discuss the uncertainties in the modelling of the physical processes and other caveats of our simulations below. After discussing the limitations in our simulations, we compare to other work and finally we discuss the implications of our results.

\subsection{Angular momentum loss}
\label{sec:dis_amloss_eda1}

As discussed in Sec. \ref{sec:consistency_eda1}, the angular momentum loss due to ablation of the disc is strongly dependent on the artificial viscosity parameter $\alpha_{\rm SPH}$ in the disc and the ablation rate of the disc also shows a decreasing trend as a function of the resolution (see Sec. \ref{sec:convergence_eda1}). Furthermore, in our simulations with Fi angular momentum is predominantly lost to the ISM through angular momentum exchange at the outer edge of the disc, instead of being carried away by stripped disc material. The angular momentum loss associated with these processes shows a decreasing trend with increasing resolution. This suggests that the ablation in our simulation, and the associated angular momentum loss, is dominated by numerical effects and that physically it may not be the dominant process for the loss of angular momentum from the disc. In the simulation of M09 with the same version of Gadget2, but with a higher resolution and smaller flow velocity (i.e. lower ram pressure), the stripping of the outer edge does not play a significant role (see Sec. \ref{sec:dis_m09_eda1}). 

If the ablation of the disc is predominantly numerical, then the time scale on which the disc loses its mass can not be considered a reliable indicator for the lifetime of the disc. Although the time scale of angular momentum loss is generally shorter than the time scale derived from the mass loss of the disc, we consider the former to be a more reliable estimator for the disc life time. In particular, we consider the time scale of angular momentum loss determined with Fi to be more indicative, because in those simulations the viscosity in the disc agrees better with accretion disc theory and the disc actually gains mass. The angular momentum loss time scale in the F16 simulation is similar to that in the G128 simulation, i.e. highest resolution with Gadget2, but it is still an order of magnitude smaller than the physical estimate in Sec. \ref{sec:visc_eda1}. 

We can not directly interpret the time scale of angular momentum loss as the lifetime of the disc, since it is dominated by numerical effects in our simulations. Physically, the decrease in specific angular momentum of the disc due to accretion of ISM with zero azimuthal angular momentum may contribute equally to, or even dominate, the decreasing disc size. Both processes are also non-linear and we consider our estimate of the angular momentum loss time scale as a lower limit to the actual lifetime of the disc. 

\subsection{Viscous evolution}

We have used two SPH codes that model the viscosity in a different way. It is clear that the transport of mass and angular momentum in the disc behaves differently in both codes even if the same viscosity prescription is implemented in both codes. To model the (viscous) evolution of the disc correctly, additional relevant physical processes, e.g. magnetohydrodynamics, radiative transfer, radial mixing (which are not all physically well understood and beyond the scope of this paper) should be incorporated. In this work, we are interested in how much mass a protoplanetary disc can sweep up and how this process would affect its lifetime. The result for the ISM loading appears to be independent of the viscous modelling of the disc and we therefore conclude that this result is robust. Furthermore, the simulations indicate that the process of face-on accretion onto a protoplanetary disc will likely shorten rather than prolong its lifetime.

\subsection{Resolution}

As mentioned in Sec. \ref{sec:set-up_eda1}, we try to find a balance between computing time and convergence. Our reference model, G16, took 3.5 days to run on 32 cores, while the comparison F16 simulation took a little more than two months on 22 CPUs. When looking at the ablation rate as a function of increasing resolution, the number of particles that are stripped increases, but the mass that they carry away decreases more rapidly. We therefore argue that the ablation rate is not caused by numerical noise but rather, as discussed above, by the treatment of viscosity in the disc. This is supported by the result that the Poisson noise for the low-resolution models that we have performed multiple times is less than 5\,\%, see table \ref{tb:gadget_eda1}. Furthermore, the ISM loading rate is robust for all our simulations and does not seem to be affected by numerical noise.

\subsection{Modelling of the ISM}

In order to be able to discern the relevant physical processes from the simulations, we have modelled the ISM in a very idealized way: as a homogeneous gas with no clumps and no turbulence. To a certain extent we have modelled the reaction of the disc to a density gradient when the flow first hits the disc. In that case the size of the disc is determined by the ram pressure. If the disc were to encounter a region of gas with lower density, the surface area of the disc can extend to larger radii without being stripped and that would probably increase its effective cross section. Despite the lower density, the ISM loading rate could still be of the same order, because a lower ISM density allows a larger surface area of the disc to collect ISM. In that sense, accretion onto protoplanetary discs may be a self-regulating process. In a follow-up work we will perform simulations with different ISM densities and velocities to investigate if and how the ISM loading rate depends on these quantities.

In principle it would be possible to add angular momentum to the disc, such that it maintains or prolongs its lifetime, if the ISM has turbulence or substructures on scales at or slightly smaller than the disc scale. However, even in that case the net effect is unlikely to increase the overall lifetime, since the disc would quickly encounter a different part of the ISM where an adverse, disruptive configuration of substructure might exist.

\subsection{Comparison to other work}
\label{sec:dis_m09_eda1}

A similar simulation has been performed by M09, but for a higher disc mass and lower velocity of the ISM with respect to the disc. In their work both the Bondi-Hoyle radius and the radius beyond which ram pressure would dominate are much larger than the radius of the disc. They used a higher resolution for the disc, initially $\sim 2.5 \times 10^5$ particles, which had 8 times the mass of their ISM particles. However, in their work the ISM also does not reside in the outer regions of the disc (see their Fig. 3), meaning that even at much higher resolution the effective surface area of the disc is smaller than its actual surface area. We interpret this as a physical effect: at the outer edge of the disc, the ISM is forced to flow around it and is not entrained. 

Furthermore, the disc in the simulation of M09 also shows a decreasing radius and steepening surface density profile as a function of time. They attribute this to the redistribution, i.e. inward movement, of disc material due to the accretion of ISM with no angular momentum. The loss of disc material to the ISM in their simulation is negligible, the disc mass only decreases due to accretion onto the star. This agrees with the trend suggested in our convergence study, i.e. that the continuous stripping of the disc at the outer edge is a numerical artefact. However, it could also be due to the much lower velocity in their work. 

A future parameter study at other, e.g. lower, densities and velocities of the ISM could provide a more decisive answer.

\subsection{Implications for the early disc accretion scenario}

Although the ISM loading rate we find corresponds relatively well to the geometric rate used in B13 for their early disc accretion scenario, they assume a disc radius that is significantly larger, i.e. 100 AU, than the radius found in our simulation. Moreover, the size of the disc decreases continuously during the process of accretion and we have assumed the idealized case in which the disc is positioned exactly perpendicular to the flow. The ISM loading rate derived in this work is therefore most probably an upper limit for the average rate one would expect for a population of discs that all have different inclinations between their rotational axis and velocity vector.

Furthermore, B13 assume a disc lifetime of $10^7$ years from observations, while we find that angular momentum transport plays a significant role in shortening the lifetime of the disc in dense ISM environments. It therefore seems unlikely that a low-mass star can accrete of the order of its own mass via its protoplanetary disc, as required in the early disc accretion scenario.

\subsection{Implications for planet formation}

The process of accretion onto protoplanetary discs may also affect planet formation. It could play a role in the formation of `hot Jupiters', i.e. massive planets that have formed further out in the protoplanetary disc and migrated inwards. The mass loading and consequent redistribution may also increase the probability of forming a planet in the habitable zone. \citet{ronco14} found that a steeper surface density profile, i.e. more mass at smaller radii, increases the probability of forming a planet with a significant water content in the habitable zone. The constraint for an initial steep surface density profile could perhaps be eased, when the entraining of ISM onto the protoplanetary disc causes the disc material to migrate inward. These suggestions could be tested by incorporating our findings in detailed planet formation models.

\section{Conclusions}

We have performed simulations of accretion of interstellar material (ISM) onto a protoplanetary disc with two different smoothed particle hydrodynamics codes. We find that, as theoretically expected, when the flow of ISM first hits the disc, all disc material beyond the radius where ram pressure dominates is stripped. As ISM is being accreted and disc material migrates inwards, the disc becomes more compact and the surface density profile increases at smaller radii. We find that the ISM loading rate, i.e. the rate at which ISM is entrained by the disc and star, is approximately constant across all our simulations with both codes and is a factor of two lower than the rate expected from geometric arguments (see Eq. \ref{eq:mdot_eda1}). This difference arises because the outskirts of the disc do not entrain ISM and therefore the effective cross section of the disc is smaller than its physical surface area. We find that, despite the accretion of ISM, the net effect is that the disc loses mass, except in the highest-resolution simulations with both codes where the disc gains mass. This decreasing trend with resolution implies that the net mass loss from the disc in our low-resolution simulations is numerical. Considering the time scale on which the disc loses all of its angular momentum rather than its mass, provides an estimate of about $10^4$ years. The angular momentum loss from the disc in our simulations is dominated by continuous stripping of disc material and by transfer of angular momentum to the ISM as it flows past the outer edge of the disc. Our convergence and consistency study as well as previous work indicate that this these effects are predominantly numerical. The time scale estimated from the simulations with the highest resolution therefore provide a lower limit to the lifetime of the disc. The loss of angular momentum due to accretion of disc material onto the star, which is governed by the (modelling of) viscous processes in the disc, is two orders of magnitude smaller than the loss associated with stripping. 

Even if the disc grows in mass, the (specific) angular momentum of the disc will always decrease in this scenario, if not for the aforementioned angular momentum loss processes then by accretion of ISM with no azimuthal angular momentum. Either way, the disc will shrink in size, thereby decreasing its effective cross section. Although our ISM loading rate agrees within a factor of two with the geometrically estimated rate, the lifetime and size of the disc are probably not sufficient to accrete the amount of mass required in the early disc accretion scenario.

In future work we will extend our simulations to explore the parameter-space and conditions that correspond to a broader range of stellar environments in order to find a parametrization for the mass loading rate onto a protoplanetary disc system in terms of the density and velocity of the ambient medium and the size of the disc. 

\begin{acknowledgements}
We thank Selma de Mink, Nate Bastian and Nickolas Moeckel for valuable discussions and input. We are also grateful to the referee whose valuable comments have improved this work. 
This research is funded by the Netherlands Organisation for Scientific Research (NWO) under grant 614.001.202. NWO also granted computational resources on Cartesius under grant SH-295-14.
\end{acknowledgements}

\bibliographystyle{aa} 
\bibliography{phdbib}

\end{document}